\documentclass[12pt]{JHEP3}
\usepackage{amsmath,amsfonts,amssymb,dsfont,ifthen,epsfig,slashed, cancel}

\newcommand{\be}{\begin{equation}}
\newcommand{\ee}{\end{equation}}
\newcommand{\ba}{\begin{eqnarray}}
\newcommand{\ea}{\end{eqnarray}}
\newcommand{\bas}{\begin{eqnarray*}}
\newcommand{\eas}{\end{eqnarray*}}
\newcommand{\nn}{\nonumber}
\newcommand\T{\rule{0pt}{2.6ex}}
\newcommand\B{\rule[-1.2ex]{0pt}{0pt}}
\newcommand{\comment}[1]{}

\newboolean{Notes}\setboolean{Notes}{true}
\newcommand{\notes}[1]
            {\ifthenelse{\boolean{Notes}}{{\tt #1}}{}}

\preprint{arXiv:0709.3813 [hep-th]}

\title{Type IIB Flows with
$\mathcal{N}=1$ Supersymmetry}
\author{Girma Hailu\thanks{hailu@lepp.cornell.edu}
\\
Newman Laboratory for Elementary Particle Physics\\
Cornell University \\
Ithaca, NY 14853 }

\abstract{\\

We write general and explicit equations which solve the supersymmetry transformations with two arbitrary complex-proportional Weyl spinors on $\mathcal{N}=1$ supersymmetric type IIB strings backgrounds with all R-R $F_1$, $F_3$, $F_5$ and NS-NS $H_3$ fluxes turned on using $SU(3)$ structures.
The equations are generalizations of the ones found for specific relations between the two spinors by Grana, Minasian, Petrini and Tomasiello in \cite{Grana:2004bg} and by Butti, Grana, Minasian, Petrini and Zaffaroni in \cite{Butti:2004pk}.
The general equations allow to study systematically generic type IIB backgrounds with $\mathcal{N}=1$ supersymmetry.
We then explore some specific classes of flows with constant axion, flows with constant dilaton, flows on conformally Calabi-Yau backgrounds, flows with imaginary self-dual 3-form flux, flows with constant ratio of the two spinors, the corresponding equations  are written down and some of their features and relations are discussed.
}

\begin{document}

\section{Introduction\label{intr}}

The gauge/gravity duality \cite{Maldacena:1998re,Gubser:1998bc,Witten:1998qj} together with flux compactification in string theories and powerful tools in supersymmetric gauge theories provide crucial insight into both the background geometries of strings and the nonperturbative dynamics of gauge theories.
In this note, we write down general and explicit equations which solve the supersymmetry transformations and which allow to study systematically $\mathcal{N}=1$ supersymmetric type IIB strings backgrounds with all $F_5$, $F_3$, $F_1$ and $H_3$ fluxes turned on using $SU(3)$ structures. We then explore the backgrounds for possible types of flows in relation to the components of the fluxes.
This is useful in scanning the backgrounds for new supergravity solutions and dual gauge/gravity theories, in studying flux compactifications with hierarchy of scales  \cite{Giddings:2001yu} and with stabilized moduli {\cite{Kachru:2003aw}, in constructing suitable cosmological models \cite{Dvali:1998pa,Kachru:2003sx} which might allow to probe stringy signatures left over from the early universe (see \cite{Tye:2006uv} for a review for instance), in studying mirror symmetry in flux compactifications (see \cite{Vafa:2000wi,Gurrieri:2002wz,Grana:2004bg} for instance), and in looking for stable flux vacua which break supersymmetry. The equations we present could be used to study specific flows on type IIB backgrounds with $\mathcal{N}=1$ supersymmetry with appropriate metric and flux ansatz.

Compactifications of type II strings over Calabi-Yau manifolds preserve $\mathcal{N}=2$ supersymmetry. On the other hand, gauge/gravity theories which exhibit physically interesting  phenomena such as nonconformal renormalization group flow, confinement and chiral symmetry breaking have reduced $\mathcal{N}=1$ supersymmetry. Such theories can be engineered by adding D-branes which turn on fluxes on Calabi-Yau backgrounds. Once fluxes are turned on, the background geometry backreacts and develops torsion and the compactification generically becomes non-Calabi-Yau.
One of the difficulties in dual gravity approaches to study QCD or when looking for supergravity flows suitable for modeling cosmological scenarios is a lack of good understanding of strings on such backgrounds. For early work on strings with torsion, see \cite{Strominger:1986uh, deWit:1986xg}. In the last five years, group structures have been used to deal with supergravity backgrounds with torsion which preserve $\mathcal{N}=1$ and $\mathcal{N}=2$ supersymmetries. See \cite{Gauntlett:2002sc, Gurrieri:2002wz, LopesCardoso:2002hd, Gauntlett:2003cy, Grana:2004bg, Butti:2004pk, Grana:2005jc,Grana:2006kf} for instance.  When the extra six dimensional space of type II strings is compactified over a generalized
Calabi-Yau with group $G$-structures, the components of the torsion and the fluxes
fall in representations of $G$.
The supersymmetry conditions translate into constraints on the balance among the components of the fluxes, the torsion, the running of the dilaton and the warp factor decomposed and organized representation by representation such that the appropriate number of supersymmetries is preserved. The equations which solve the constraints can be efficiently used to study the backgrounds.

Type IIB theory has two Majorana-Weyl spinors of the same chirality. In order to preserve $\mathcal{N}=1$ supersymmetry, only one covariantly constant spinor is necessary. In flux compactifications with $SU(3)$ structures, there is one globally defined $SU(3)$ singlet spinor on the extra 6-d space which can be arranged to be covariantly constant with respect to a Levi-Civita connection containing torsion and the fluxes included.\footnote{The 4-d spacetime is taken to be conformally flat with a warp factor which is a function of the coordinates on the extra 6-d space.} Let us denote the positive chirality component of the $SU(3)$ singlet spinor in 6-d by $\eta_+$. Let us also denote the positive chirality 6-d components of the two Majorana-Weyl 10-d spinors by $\eta_+^{1}$ and $\eta_+^{2}$.
The two spinors $\eta_+^{1}$ and $\eta_+^{2}$ are then complex-proportional to the globally defined singlet spinor $\eta_+$ and to each other on backgrounds with $\mathcal{N}=1$ supersymmetry. The relations between these spinors can be expressed
in terms of two complex parameters $\alpha$ and $\beta$ which are functions of the coordinates on the extra 6-d space as
\be \eta_+^1=\frac{1}{2}(\alpha+\beta) \eta_+,\qquad
\eta_+^2=\frac{1}{2i}(\alpha-\beta) \eta_+.\label{eta1eta2-def}\ee
The parameters $\alpha$ and $\beta$ are tied to the components of the fluxes which are turned on, the dilaton, the warp factor, and the torsion while $\mathcal{N}=1$ supersymmetry is preserved. The cases of $\alpha=0$, $\beta=0$, $\beta=\pm \alpha$ or $\beta=\pm i\alpha$ are special because the supersymmetry transformations simplify considerably and have been extensively studied, see \cite{Dall'Agata:2004dk,Grana:2004bg} for instance for discussion.

Here, we present the general equations which solve the supersymmetry transformations and which accommodate complex-proportional Weyl spinors with arbitrary magnitudes and phase between them. The equations generalize the ones found for specific relations between the spinors by Grana, Minasian, Petrini and Tomasiello (GMPT) in \cite{Grana:2004bg} and by Butti, Grana, Minasian, Petrini, and Zaffaroni (BGMPZ) in \cite{Butti:2004pk}.
The decomposition of the equations falls in $1\oplus1$, $8\oplus8$, $6\oplus \bar{6}$ and $3\oplus\bar{3}$ representations of $SU(3)$. The ones in the $1\oplus1$, the $8\oplus8$ and the $6\oplus\bar{6}$ sectors are the same as in \cite{Grana:2004bg}. The $3\oplus\bar{3}$ sector involves the equations which govern the running of the dilaton, the warp factor and all the fluxes and that is where we present new and general equations which together with the equations in the $1\oplus1$, the $8\oplus8$ and the $6\oplus\bar{6}$ sectors allow to scan the moduli space of $IIB$ backgrounds with $SU(3)$ structures and $\mathcal{N}=1$ supersymmetry.
We reproduce the equations obtained with a gauge choice in \cite{Grana:2004bg} as a specific case of the general equations with a constant phase of zero between $\alpha$ and $\beta$, or equivalently a constant phase of $\pi/2$ between the two spinors $\eta_+^1$ and $\eta_+^2$.
Figure \ref{fig-albe} schematically shows the $(\alpha,\beta)$ spinors parameters space.
\begin{figure}[ht]
\begin{center}
\leavevmode
\includegraphics[width=0.3\textwidth, angle=0]{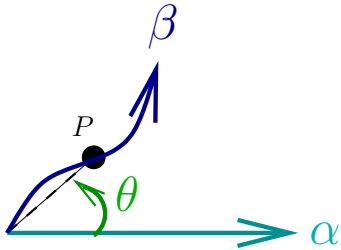}
\caption[]{Spinors parameters space. The parameters $\alpha$ and $\beta$ define the relations between the two complex-proportional Weyl spinors and the $SU(3)$ invariant spinor. The curved line with both the magnitude and the phase of $\beta$ changing and the horizontal line with the magnitude of $\alpha$ changing show an example of a generic flow in the spinors parameters space. The phase of $\alpha$ is taken to be constant in this figure. The angle $\theta$ shows the phase between $\alpha$ and $\beta$ and the dashed line shows the magnitude of $\beta$ for the point $P$.
The magnitudes of $\alpha$, $\beta$ and $\theta$ are tied to the components of the fluxes, the torsion, the running of the dilaton and the warp factor.  The equations we present here could be used to study flows systematically in the whole parameters space. }
\label{fig-albe}
\end{center}
\end{figure}

Our initial motivation for investigating these backgrounds was because corrections to the anomalous mass dimension on the gauge theory side in the Klebanov-Strassler throat \cite{Klebanov:2000hb}, where type $IIB$ string theory with $N$ D3- and $M$ D5-branes on $AdS_5\times T^{1,1}$ background is dual to $\mathcal{N}=1$ supersymmetric $SU(N+M)\times SU(N)$ gauge theory with bifundamental chiral superfields and a quartic tree level superpotential in four dimensions, lead to
supergravity flows with running dilaton and generic relations between the two spinors or require turning on the $F_1$ flux \cite{Hailu:2006uj}.

The organization of this article is as follows. We start with considering backgrounds with fluxes and metric which preserve 4-d Poincare invariance and a review of the supersymmetry transformations and the decomposition of the fluxes and the torsion in $SU(3)$ representations in \cite{Grana:2004bg}.
We then present the general equations which solve the supersymmetry transformations in the whole $(\alpha,\beta)$ parameters space and with all fluxes turned on. The independent set of relations will be explicitly written for the fluxes, the torsion, and the running of the dilaton and the warp factor in terms of the spinors parameters.
We continue with considering some specific classes of flows in which the backgrounds are conformally Calabi-Yau, flows with imaginary self-dual 3-form flux, flows with constant axion, flows with constant dilaton, flows with constant ratio of the two spinors and the corresponding equations are written down.

Relations among the fluxes, the dilaton, the warp factor and the background geometry, some of which are familiar, follow and are proved using the equations. For instance,
the flux and the torsion components in the singlet representation (the ones in the $(0,3)$ and $(3,0)$ forms) cannot balance each other and the singlet components of the 3-form fluxes must vanish identically.
Flows with constant axion, constant dilaton and nonconstant warp factor are conformally Calabi-Yau.
Conformally Calabi-Yau flows have imaginary self-dual 3-form flux.
Flows with imaginary self-dual 3-form flux have primitive 3-form fluxes.
Flows with imaginary self-dual 3-form flux have constant dilaton-axion coupling coefficient $\tau=ie^{-\Phi}+C_0$.
Therefore, for instance,
flows with imaginary self-dual 3-form flux and constant axion have constant dilaton.
Flows with imaginary self-dual 3-form flux and constant dilaton have constant axion.
Conformally Calabi-Yau flows with constant axion have constant dilaton.
Flows with constant axion and nonconstant dilaton have nonprimitive 3-form flux.
Flows with constant dilaton and nonconstant axion have nonprimitive 3-form flux.

The independent equations are summarized in Appendix \ref{appA}.

\section{Supersymmetry transformations}

In this section, we make a brief review and write the supersymmetry transformations in terms of component fields in $SU(3)$ representations obtained in \cite{Grana:2004bg}.
The backgrounds we want to study are type IIB with
the dilaton $\Phi$, NS-NS 2-form potential $B_2$ and the corresponding 3-form flux $H_3=dB_2$, R-R 0-, 2- and 4-form potentials $C_0$, $C_2$ and $C_4$  \cite{Polchinski:1998rr}. Let us define the (modified) R-R fluxes,
\be
{F}_1= d C_0,
\quad {F}_3= d C_2-C_0 H_3,\quad F_5=d C_4-H_3\wedge C_2,\ee
where we need to impose that the 5-form flux be self-dual and write ${\tilde{F}}_5= (1+\star)F_5$.
Type IIB theory contains two fermions: a gravitino which is denoted by $\psi$ and a dilatino which is denoted by $\lambda$. We work with type IIB action in string frame with the string coupling absorbed in the dilaton.

For compactifications preserving four dimensional Poincare invariance such that $R^{(1,9)} \to R^{(1,3)}\times Y$ and $Spin(1,9)\to Spin(1,3)\otimes Spin(6)$, we can write the 10-d metric as
\be
ds_{10}^2=e^{2A{(y)}}dx_\mu dx^\mu +ds_6^2(y),\label{metricg10d}
\ee
where $x$ denotes the 4-d coordinates on $R^{(1,3)}$, $y$ denotes the coordinates on the extra 6-d manifold $Y$, and $A(y)$ is the warp factor which depends only on the coordinates in the extra space.
Our notation is such that uppercase indices $M$, $N$, $\cdots$ run over all the 10-d coordinates, lowercase indices $m$, $n$, $\cdots$ run over the extra 6-d space coordinates and take values 1 to 6, and the indices $\mu$, $\nu$, $\cdots$ run over the 4-d spacetime coordinates.
In order to preserve 4-d Poincare invariance, all the fluxes $F_1$, $F_3$, $H_3$ and the $F_5$ part of the self-dual $\tilde{F}_5$ have only internal (extra space) components, or if we think in terms of D-branes giving rise to these fluxes via geometric transition, the D-branes fill up 4-d spacetime with remaining components wrapping cycles in the extra space.
The $32\times32$ gamma matrices which generate the Clifford algebra of $Spin(1,9)$ in 10-d are also decomposed as $\Gamma^{\mu}=\gamma^{\mu}\otimes1$ and $\Gamma^{m}=\gamma_{5}\otimes \gamma^{m}$, where $\gamma^{\mu}$ and $\gamma^{m}$ are respectively $4\times 4$ and $8\times8$ matrices which generate the algebras of $Spin(1,3)$ and $Spin(6)$ respectively.
The spinor representations in 4-d and in 6-d are respectively given by the eigenvalues of $\gamma_5\equiv-i\gamma_0\gamma_1\gamma_2\gamma_3$ and $\gamma_7\equiv i\gamma_1\gamma_2\gamma_3\gamma_4\gamma_5\gamma_6$. The positive and the negative chirality projections of a spinor $\eta$ in 6-d are written as $\eta_{\pm}=(1\pm\gamma_7)\eta/2$.

When type IIB theory is compactified over Calabi-Yau threefold, it gives $\mathcal{N}=2$ supersymmetry in 4-d. Let us denote the two 4-component Majorana spinors parameters for the $\mathcal{N}=2$ supersymmetry transformations in 4-d by $\zeta^1$ and $\zeta^2$ with their respective 2-component positive and negative chirality Weyl components $\zeta^1_{\pm}$ and $\zeta^2_{\pm}$ such that $\zeta^i_{-}={\zeta_{+}^i}^{*}$.
Let us also denote the two Majorana-Weyl spinors of the same chirality for the supersymmetry transformations in 10-d by $\epsilon^{1}$ and $\epsilon^{2}$.
The 10-d spinors can then be decomposed as
\be\epsilon^1=\zeta^1_{+}\otimes\eta_{+}+\zeta^1_{-}\otimes\eta_{-},\quad \epsilon^2=\zeta^2_{+}\otimes\eta_{+}+\zeta^2_{-}\otimes\eta_{-},\label{spinordec1}\ee where $\eta_{\pm}$ are 4-component Weyl spinors on $Y$  such that $\eta_{-}={\eta_{+}}^{*}$.
Our interest is in backgrounds which preserve only $\mathcal{N}=1$ supersymmetry in four dimensions. On such backgrounds, there is only one independent spinor in 4-d. Let us denote the positive chirality component of this 4-d spinor by $\zeta_+$. The two positive chirality spinors $\zeta_+^1$ and $\zeta_+^2$ are then complex-proportional to $\zeta_+$ and to each other on backgrounds with $\mathcal{N}=1$ supersymmetry. Let us write the relations as
\be \zeta_+^1=\frac{1}{2}(\alpha+\beta) \zeta_+,\qquad
\zeta_+^2=\frac{1}{2i}(\alpha-\beta) \zeta_+,\label{zeta1eta2-def}\ee
where $\alpha$ and $\beta$ are complex parameters.
Because we take the 4-d spacetime to be flat except for the warp factor $e^{2A(y)}$ in the metric given by (\ref{metricg10d}), the spinor $\zeta_+$ is constant with $\alpha$ and $\beta$ being functions of the coordinates on the extra space. We can rewrite the decomposition (\ref{spinordec1}) with the complex-proportionality coefficients absorbed in the spinors on $Y$ as
\be\epsilon^1=\zeta_{+}\otimes\eta^1_{+}+\zeta_{-}\otimes\eta^1_{-},\quad \epsilon^2=\zeta_{+}\otimes\eta^2_{+}+\zeta_{-}\otimes\eta^2_{-},\label{spinordec2}\ee
where $\eta^{i}_{-}={\eta^{i}_{+}}^{*}$.  The spinors $\eta_+^{1}$, $\eta_+^2$ and $\eta_+$ are then related as given by (\ref{eta1eta2-def}), $ \eta_+^1=\frac{1}{2}(\alpha+\beta) \eta_+$ and
$\eta_+^2=\frac{1}{2i}(\alpha-\beta) \eta_+$.

But a generic compactification of $Y$ with structure group $SO(6)\sim SU(4)$ has no globally defined covariantly constant spinor and gives no supersymmetry. The spinor representation of $SO(6)$
corresponds to the fundamental
representation of $SU(4)$ which decomposes as $1\oplus3$ under $SU(3)$.
Thus there is one $SU(3)$ singlet spinor on $Y$.
In order to preserve some supersymmetry, $Y$ needs to have a reduced structure group and to preserve $\mathcal{N}=1$ supersymmetry the structure group on $Y$ has to be reduced at least to $SU(3)$. In that case, we can take the $\eta_{+}$ discussed above (and which appears in (\ref{eta1eta2-def})) to be this $SU(3)$ singlet spinor. If $Y$ were a Calabi-Yau threefold, then the manifold would have $SU(3)$ holonomy and give $\mathcal{N}=2$ supersymmetry in four dimensions, and the globally invariant spinor would be covariantly constant. However, on  $\mathcal{N}=1$ supersymmetric backgrounds with $SU(3)$ structures with fluxes turned on, the globally invariant spinor $\eta_+$ is not covariantly constant with respect to the usual Levi-Civita connection but with a connection which includes torsion and the fluxes included.

The supersymmetry transformations of the gravitino and the dilatino fields in 10-d can be expressed as, see \cite{Bergshoeff:2001pv} for instance,
\be\label{st-psi}
\delta \psi_M = \nabla_M\,\epsilon -\frac{1}{4} (\slashed{H}_3)_{M}\sigma^3\,\epsilon+\frac{1}{8}e^{\Phi}\Bigl( \slashed{F}_{3}\sigma^1
+i(\slashed{F}_{1}+\slashed{F}_{5})\sigma^2\Bigr)\Gamma_M\,\epsilon,
\ee
\be\label{st-lam}
\delta \lambda=\slashed{\partial}\Phi\,\epsilon-\frac{1}{2}\slashed{H}_3\sigma^3\,\epsilon
-\frac{1}{2}e^{\Phi} \Bigl(\slashed{F}_{3}\sigma^1+2i\slashed{F}_{1}\sigma^2\Bigr)\,\epsilon,
\ee
where $\sigma^i$ are the $2\times 2$ Pauli matrices which now act on the supersymmetry transformation parameter $\epsilon$ with the two Majorana-Weyl spinors as components, \be\epsilon=\left(
                  \begin{array}{c}
                    \epsilon^1 \\
                    \epsilon^2 \\
                  \end{array}
                \right)=\left(
                  \begin{array}{c}
                    \zeta_{+}\otimes\eta^{1}_{+}+\zeta_{-}\otimes\eta^{1}_{-} \\
                    \zeta_{+}\otimes\eta^{2}_{+}+\zeta_{-}\otimes\eta^{2}_{-} \\
                  \end{array}
                \right).\label{spinor10dee}\ee
The slash is for contraction with gamma matrices with the definition for contraction of
$p-q$ number of components of a $p$-form,
\be
(\slashed{\omega}_p)_{M_{1}\cdots M_q}= \frac{1}{(p-q)!}(\omega_p)_{M_{1}\cdots M_qM_{q+1}\cdots M_p}\Gamma^{[M_{q+1}}\cdots \Gamma^{M_p]}
\ee
and $\nabla_M$ is the covariant derivative.
The combination $\Gamma^M \delta\psi_M-\delta\lambda$ contains no R-R flux and the two supersymmetry transformations given by (\ref{st-psi}) and (\ref{st-lam}) can be traded for (\ref{st-psi}) and
\be\label{st-lam2}
\Gamma^M \delta\psi_M-\delta\lambda=\slashed{\nabla}\,\epsilon -\slashed{\partial}\Phi\,\epsilon-\frac{1}{4}\slashed{H}_3\sigma^3\,\epsilon.
\ee

Let us start with parameterizing the metric on $Y$ as
\be\label{ds6s-1}
ds_6^2=\delta_{mn}G^m G^n,
\ee
where $G^m$ are real differential 1-forms which are expressed in terms of linear combinations of the coordinate 1-forms $dy^n$ on $Y$ with coefficients which are functions of $y$.
We also define the $\gamma$ matrices in 6-d with respect to this parametrization of the metric such that \be\{\gamma^m,\gamma^n\}=2\delta^{mn}.\ee
The supersymmetry transformations in terms of flux and torsion components in $SU(3)$ representations are conveniently written in complex basis.
Let us define the complex 1-forms\footnote{The final expressions at the end of this section are for the coefficients in the supersymmetry transformations expanded in terms of these complex forms.}
\be\label{ZiGi}
Z^1=G^1+i G^2,\quad Z^2=G^3+i G^4,\quad Z^3=G^5+i G^6
\ee
and their complex conjugates $\bar{Z}^{\bar{i}}=({Z}^{i})^\ast$.
(The holomorphic/antiholomorphic indices $i$ and $\bar{i}$ run over 1 to 3.) Here $G^m$ and $Z^i$ are not closed.  Demanding that the backgrounds preserve supersymmetry leads to constraints which make $Y$ a complex manifold.  Rewriting (\ref{ds6s-1}),
\be
ds_6^2=\delta_{i\,\bar{j}}Z^{i}\bar{Z}^{\bar{j}}.
\ee

We also define $\gamma$ matrices with holomorphic/antiholomorphic indices $\gamma^i$ and $\bar{\gamma}^{\bar{i}}$ in terms of $\gamma^m$
as
\be
\gamma^{i}=(\gamma^{2m-1}+i\gamma^{2m})\delta^i_m,\quad \bar{\gamma}^{\bar{i}}=(\gamma^{2m-1}-i\gamma^{2m})\delta^{\bar{i}}_m.
\ee
Note that
\be\{\gamma^i,\bar{\gamma}^{\bar{j}}\}=4\delta^{i\bar{j}}, \quad\{\gamma^i,\gamma^j\}=0, \quad \{\bar{\gamma}^{\bar{i}},\bar{\gamma}^{\bar{j}}\}=0.\ee

The supersymmetry variation (\ref{st-psi}) for $M=\mu$ gives $\Gamma^{\mu}\delta \psi_\mu=0$ and, with the metric (\ref{metricg10d}) and recalling that the fluxes have only internal components,
\be
\slashed{\partial}A\,\epsilon-\frac{1}{4}e^{\Phi}(\slashed{F}_{3}\sigma^1
+i(\slashed{F}_{1}+\slashed{F}_{5})\sigma^2)\,\epsilon=0.\label{st-psi-2}
\ee
Using (\ref{spinor10dee}) and (\ref{eta1eta2-def}), (\ref{st-psi-2}) can be written in terms of the $SU(3)$ invariant spinor on $Y$ as
\be\label{susypsimu1a}
\alpha\slashed{\partial}A\, \eta_{+}-\frac{i}{4}e^{\Phi}(\beta \slashed{F}_3-\alpha (\slashed{F}_{1}+ \slashed{F}_5))\eta_{+}=0,
\ee
\be\label{susypsimu2a}
\beta\slashed{\partial}A \,\eta_{+}+\frac{i}{4}e^{\Phi}(\alpha \slashed{F}_3-\beta (\slashed{F}_{1}+\slashed{F}_5))\eta_{+}=0,
\ee
where the slashes are now for contractions with the $\gamma$ matrices in 6-d.
Note that the second equation (\ref{susypsimu2a}) can be obtained from the first (\ref{susypsimu1a}) by interchanging $\alpha\leftrightarrow \beta$ and flipping the signs of the R-R fluxes and this is the case throughout and, therefore, we will write only one of such equations in the remaining part of this section.
The variation (\ref{st-psi}) for $M=m$ gives
\be\label{susypsim1}
(\alpha \nabla_{m}+\partial_{m} \alpha -\frac{1}{4}\beta (\slashed{H}_3)_{m})\eta_{+}+\frac{i}{8}e^{\Phi}(\beta \slashed{F}_3-\alpha (\slashed{F}_{1}+\slashed{F}_5))\gamma_{m}\eta_{+}=0.
\ee
Moreover, (\ref{st-lam2}) gives
\be
\alpha(\slashed{\nabla}+2\slashed{\partial}A+\slashed{\partial}\ln {\alpha}-\slashed{\partial}\Phi)\eta_{+}-\frac{1}{4}\beta \slashed{H}_3\eta_{+}=0.\label{susylam}\ee

The complex and K\"{a}hler structures on $Y$ are determined by properties in the variations of the fundamental 2-form, denoted by $J$, and the holomorphic 3-form, denoted by $\Omega$.  Let us summarize the decompositions of the variations of $J$ and $\Omega$ with components in representations of $SU(3)$ when $Y$ has $SU(3)$ structures. See \cite{Chiossi:2002tw} for instance for details.
We write $J$ and $\Omega$ as
\begin{eqnarray}\label{JZ-1}
J&=& \frac{1}{2}J_{i\bar{j}}\,{Z^{i}}{ {\wedge} }{\bar{Z}^{\bar{j}}}=
\frac{i}{2} \delta_{i\bar{j}}{Z^{i}}{ {\wedge} }{\bar{Z}^{\bar{j}}},
\end{eqnarray}
\begin{eqnarray}\label{OmZ-1}
\Omega &=&\frac{1}{6}\Omega_{ijk}{Z^i}{ {\wedge} }{Z^j}{ {\wedge} }{Z^k}=\frac{1}{6}\epsilon_{ijk}{Z^i}{ {\wedge} }{Z^j}{ {\wedge} }{Z^k}=
{Z^1}{ {\wedge} }{Z^2}{ {\wedge} }{Z^3}.
\end{eqnarray}
Note that $J$ is a $(1,1)$ form and $\Omega$ is a $(3,0)$ form with respect to the holomorphic/antiholomorphic 1-forms.
The variation $dJ$ has components with $(2,1)\oplus (1,2)\oplus(3,0)\oplus(0,3)$ forms. Moreover, $d\Omega$ has components with $(3,1)\oplus (2,2)$ forms; it does not have a $(4,0)$, since a complex 4-form vanishes in three complex dimensions. For $Y$ with $SU(3)$ structures, the components can be broken down in representations of $SU(3)$. The $(3,0)\oplus(0,3)$ forms in $dJ$ give components in the singlet representation. The $(2,1)\oplus (1,2)$ forms give components in the $(\bar{6}\oplus3) \oplus (6\oplus\bar{3})$ representations, where the components in the $\bar{6}$ and the $6$ representations are the primitive parts (have no components of the form $\cdots\wedge J$ and contracting them with $J$ gives zero). Our notation is such that the components in the $6\oplus\bar{3}$ representations in the 3-forms come from $(1,2)$ forms. The $(3,1)$ form in $d\Omega$ gives component in the $\bar{3}$ representation and the $(2,2)$ form gives components in the $8\oplus1$ representations. All in all, the variations can be written as
\be\label{dJ-1a}
dJ=-\frac{3}{2}\mathrm{Im}(W_1^{(1)} \bar{\Omega})+(W_4^{(3)}+W_4^{(\bar{3})})\wedge J+(W_3^{(6)}+W_3^{(\bar{6})}),
\ee
\be\label{dOm-1a}
d\Omega=W_1^{(1)} J^2+W_2^{(8)}\wedge J+W_5^{\bar{3}}\wedge \Omega,
\ee
where the $W$'s are the torsion components with the superscripts denoting the $SU(3)$ representations and the subscripts denoting five torsion classes.
If $Y$ is a Calabi-Yau manifold with complex structure defined by (\ref{ZiGi}), then both $J$ and $\Omega$ are closed, $dJ=0$ and $d\Omega=0$, and all torsion classes vanish. Thus nonvanishing components of the torsion measure the departure of the manifold from being Calabi-Yau. For $Y$ to be a complex manifold, the $(3,0)$ and $(0,3)$ forms in $dJ$ and the $(2,2)$ forms in $d\Omega$ need to vanish (since a variation on a complex manifold should raise the holomorphic and the antiholomorphic forms only once and separately) which implies that $W_1^{(1)}$ and $W_2^{(8)}$ vanish. These constraints actually follow from the supersymmetric conditions, without a need to be assumed a priori.

The 3-form fluxes, $F_3$ and $H_3$, have only internal components with $(0,3)$ and $(1,2)$ forms and their conjugates and can be decomposed as $dJ$,
\be\label{H3-1a}
H_3=-\frac{3}{2}\mathrm{Im}(H_3^{(1)} \bar{\Omega})+(H_3^{(3)}+H_3^{(\bar{3})})\wedge J+(H_3^{(6)}+H_3^{(\bar{6})}),
\ee
\be\label{F3-1a}
F_3=-\frac{3}{2}\mathrm{Im}(F_3^{(1)} \bar{\Omega})+(F_3^{(3)}+F_3^{(\bar{3})})\wedge J+(F_3^{(6)}+H_3^{(\bar{6})}).
\ee
The $F_5$ part of the self-dual 5-form flux has only internal components and is written as
\be
{F_5}=(F_5^{(3)}+F_5^{(\bar{3})})\wedge J\wedge J.
\ee

The supersymmetry transformations given by (\ref{susypsim1}) and (\ref{susylam}) contain the covariant derivative of the spinor which needs to be expressed in terms of torsion components in $SU(3)$ representations in order for both the flux and the torsion components to come in symmetrically.
The covariant derivative of $\eta$ could  be written, with the spinor normalized to constant,  as
\be\label{qdef-1} \nabla_{m}\eta=i(q_{m}\gamma_7+q_{mn}\gamma^{n})\eta.
\ee
Note that, in terms of holomorphic/antiholomorphic indices, $q_{\bar{i}}\sim \bar{3}$, $q_{ij}\sim 3\otimes{3}=6\oplus \bar{3}$ and $q_{i\bar{j}}\sim 3\otimes{\bar{3}}=8\oplus 1$. Thus $q_{\bar{i}}$, $q_{ij}$, $q_{i\bar{j}}$ and their conjugates contain the same degrees of freedom as $W_5^{(\bar{3})}$, $W_4^{(\bar{3})}$, $W_3^{(6)}$, $W_2^{(8)}$, $W_1^{(1)}$ and their conjugates.
The relations between the two sets of torsion components are given in \cite{Fidanza:2003zi, Grana:2004bg}.

The supersymmetry transformations given by (\ref{susypsimu1a})-(\ref{susylam}) are expressed in terms of the invariant spinor $\eta_+$ with standard basis in terms of products of $\gamma$ matrices. In particular, the basis $\gamma$ matrices in 6-d can be taken as
\be
1,\quad \gamma^m,\quad \gamma^{m_1m_2},\quad\cdots, \gamma^{m_1\cdots m_6}.
\ee
The supersymmetry transformations are conveniently written in a basis with coefficients which are in representations of $SU(3)$ such that the flux and the torsion components come on the same footing by projecting them to a basis with elements $\eta_{\pm}$ and $\gamma^m\eta_{\pm}$. The projection was done in \cite{Grana:2004bg}.
A useful ingredient used in doing the projection and calculating the coefficients was a characterization of the generalized complex structure in terms of pure spinors \cite{Hitchin:2004ut} where the tensor product of the standard spinors is related to the pure spinors via
\be
\eta_{\pm}\otimes \eta_{\pm}^{\dagger}=\frac{1}{8}\cancel{e^{\mp i {J}}},\quad \eta_{+}\otimes \eta_{-}^{\dagger}=-\frac{i}{8}\cancel{\Omega},\quad \eta_{-}\otimes \eta_{+}^{\dagger}=-\frac{i}{8}\cancel{\bar{\Omega}}.\label{pure-sp-def}
\ee
The $J$ and the $\Omega$ in (\ref{pure-sp-def}) are the fundamental 2-form and the holomorphic 3-form and the slash is for contraction with $\gamma^m$, now after promoting the wedge product in a $p$-form in each term to antisymmetric product of $\gamma$ matrices, \be w_{m_1m_2\cdots m_p}G^{m_1}\wedge G^{m_2}
\wedge\cdots\wedge{G^{m_p}}\to w_{m_1m_2\cdots m_p} \gamma^{m_1m_2\cdots m_p}.\label{pure-stand}\ee The slashed objects on the right hand sides in (\ref{pure-sp-def}) are the pure spinors.
The projection to the basis with components in $\eta_{\pm}$ and $\gamma^m \eta_{\pm}$ is then done by multiplying the supersymmetry variations by $\eta_{\pm}^\dagger$ and $\eta_{\pm}^\dagger\gamma^m$ from the left, using the relation between the standard spinors and the pure spinors, (\ref{pure-sp-def}), mapping the sum of forms to the sum of antisymmetric products of $\gamma$ matrices, (\ref{pure-stand}), and taking traces.

Finally, let us write the coefficients in the gravitino and the dilatino supersymmetry transformations (\ref{susypsimu1a}), (\ref{susypsim1}) and (\ref{susylam}) projected to the $\eta_{\pm}$ and $\gamma^m \eta_{\pm}$ basis, decomposed in terms of the components of the fluxes and the torsion in $SU(3)$ representations, and expressed in terms of holomorphic/antiholomorphic indices. Each coefficient needs to vanish identically and using the coefficients in the supersymmetry transformations obtained in \cite{Grana:2004bg}\footnote{Some of the terms in (\ref{Aib}) and (\ref{TibA}) have coefficients and signs different from \cite{Grana:2004bg}. However, the general  equations we obtain in the next section reproduce the equations obtained in \cite{Grana:2004bg} as a specific case of fixed phase of zero between $\alpha$ and $\beta$ with $F_1$ turned off.},
\ba\label{QijW}
 \Omega_{ij}^{\,\,\,\,\bar{k}} (\alpha W^{(\bar{3})}_4 -i \beta H_3^{(\bar 3)})_{\bar{k}}
+ \frac{i}{2} (\alpha W^{(6)}_3
- i \beta H_3^{(6)})_{\bar{k}\bar{l}i} \Omega_{j}^{\,\,\,\bar{k}\bar{l}} \nn\\ + \frac{i}{2} e^{\Phi} \Big(
  \Omega_{ij}^{\,\,\,\,\bar{k}}  (\alpha F^{(\bar{3})}_{1}
+2 \alpha F^{(\bar{3})}_{5})_{\bar{k}}-\beta (F^{(6)}_{3})_{\bar{k}\bar{l}(i} \Omega_{j)}^{\,\,\,\bar{k}\bar{l}}\Big)=0\,,
\ea
\be\label{QibWH}
\partial_{\bar{i}} \alpha +\frac{1}{2} \Bigl( \alpha (W^{({\bar{3}})}_5 - W^{(\bar{\bar{3}})}_4) -
i \beta H_3^{(\bar{3})} \Bigr)_{\bar{i}}=0\,,
\ee
\be\label{Aib}
\alpha \partial_{\bar{i}} A+\frac{i}{4} e^{\Phi}
\Big(\alpha F^{(\bar{3})}_1 - 2 i \beta F_3^{(\bar{3})}
- 2 \alpha F_5^{(\bar{3})} \Big)_{\bar{i}}=0\,,
\ee
\be\label{TibA}
\alpha \partial_{\bar{i}} (2A -\Phi + \log \alpha)
+ \frac{1}{2} \left( \alpha (W^{(\bar{3})}_4 + W^{(\bar{3})}_5) -i \beta H_3^{(\bar{3})} \right)_{\bar{i}}=0,
\ee
\be\label{QibjW}
(\alpha W^{(1)}_1 + 3 i \beta H_3^{(1)}) \eta_{\bar{i}
  j} + i \alpha (W^{(8)}_{2})_{\bar{i} j} =0\,,
\ee
\be\label{SF31}
\beta e^{\Phi}F_3^{(1)}=0
\ee
\be\label{TWH}
i\alpha W_1^{(1)} + \beta H_3^{(1)}=0,
\ee
Here we have written only the independent constraints in terms of the components in the $6\oplus \bar{3}\oplus 1\oplus 8$ representations which come from the $(1,2)$,  $(0,3)$ and $(2,2)$ forms so that the set of constraints can be solved readily. All the independent set of equations will be written down in terms of these components and we will often refer only to these forms and representations in our discussions.
There is also a second set of equations, because of the two sets of supersymmetry variations in (\ref{susypsimu1a})-(\ref{susylam}) following from the two Weyl components, which can be written down by exchanging the parameters $\alpha \leftrightarrow \beta$ and flipping the sign of the R-R fluxes in all expressions above as we discussed earlier in this section.

\section{General equations}

In this section, we write down the general equations which solve the constraints from the supersymmetry transformations given by (\ref{QijW})-(\ref{TWH}).
The ones in the $1\oplus1$, the $8\oplus8$ and the $6\oplus\bar{6}$ sectors are the same as in \cite{Grana:2004bg}. The equations in the $3\oplus\bar{3}$ sector involve the equations which govern the running of the dilaton and the warp factor in addition to the flux and the torsion components. The $3\oplus\bar{3}$ sector is subtle and that is where we present new and  general equations which together with the equations in the $1\oplus1$, the $8\oplus8$ and the $6\oplus\bar{6}$ sectors allow to scan the moduli space of $IIB$ backgrounds with $SU(3)$ structures and $\mathcal{N}=1$ supersymmetry.

\subsection{The $1\oplus1$ and the $8\oplus8$ sectors}

The equations in the  $1\oplus1$ and the $8\oplus8$ sectors follow from the constrains given by (\ref{QibjW})-(\ref{TWH}) together with the second set of constraints which can be read off with the change of variables  $\alpha \leftrightarrow \beta$ and  $F^{(1)}_{3} \rightarrow -F^{(1)}_{3}$,
\be\label{SF31b}
\beta e^{\Phi}F_3^{(1)}=0,\qquad \alpha e^{\Phi}F_3^{(1)}=0,
\ee
\be\label{QibjWb}
\alpha (W^{(8)}_{2})_{\bar{i} j} =0,\qquad \beta (W^{(8)}_{2})_{\bar{i} j} =0,
\ee
\be\label{QibjWc}
(\alpha W^{(1)}_1 + 3 i \beta H_3^{(1)}) \eta_{\bar{i}
  j} =0,\qquad (\beta W^{(1)}_1 + 3 i \alpha H_3^{(1)}) \eta_{\bar{i}
  j} =0,
\ee
\be\label{TWHb}
i\alpha W_1^{(1)} + \beta H_3^{(1)}=0,\qquad i\beta W_1^{(1)} + \alpha H_3^{(1)}=0.
\ee
Solving the above constraints, we find the relations necessary to preserve $\mathcal{N}=1$ supersymmetry with fluxes turned on (and, therefore, at least one of $\alpha$ or $\beta$ nonzero\footnote{In fact, even more, $\alpha$ and $\beta$ cannot both be constants on nontrivial $\mathcal{N}=1$ flux backgrounds, since the fundamental 2-form and the holomorphic 3-form would both then be closed (and all fluxes would vanish and the backgrounds would reduce to Calabi-Yau). More discussion about stationary points in the spinors parameters space will be given in section \ref{sec-ccy}. \label{footnotealbe0}}) in the $1\oplus1$ and the $8\oplus8$ sectors,
\be
W^{(1)}_1=0,\quad W^{(8)}_2=0,\quad H^{(1)}_3=0,\quad F^{(1)}_3=0.
\ee
Recall that the vanishing of the singlet and the octet components of the torsion was necessary in order to make $Y$ complex.  Now we see that each and every one of the singlet components of the torsion and the fluxes and the octet component of the torsion need to vanish identically. In other words, the flux and the torsion components in the singlet representation cannot be arranged to balance on backgrounds with $\mathcal{N}=1$ supersymmetry.

\subsection{The $6\oplus\bar{6}$ sector}\

In this sector, we have components of the 3-form fluxes which need to balance with the $W_3$ torsion in order to preserve $\mathcal{N}=1$ supersymmetry.  In particular, (\ref{QijW}) requires
\be\label{QR61}
(\alpha W^{(6)}_3
- i \beta H_3^{(6)})_{\bar{k}\bar{l}(i} \Omega_{j)}^{\,\,\,\bar{k}\bar{l}} =  \beta (F^{(6)}_{3})_{\bar{k}\bar{l}(i} \Omega_{j)}^{\,\,\,\bar{k}\bar{l}}
\ee
and the second constraint is obtained with the change of variables  $\alpha \leftrightarrow \beta$ and  $F^{(6)}_{3} \rightarrow -F^{(6)}_{3}$,
\be\label{QR62}
(\beta W^{(6)}_3
- i \alpha H_3^{(6)})_{\bar{k}\bar{l}(i} \Omega_{j)}^{\,\,\,\bar{k}\bar{l}} =  -\alpha (F^{(6)}_{3})_{\bar{k}\bar{l}(i} \Omega_{j)}^{\,\,\,\bar{k}\bar{l}}.
\ee
These two constraints are solved to find two independent complex equations which relate $F^{(6)}_{3}$, $H^{(6)}_{3}$ and $W^{(6)}_{3}$ and were obtained in \cite{Grana:2004bg},
\be\label{W3F6}
(\alpha^2-\beta^2)W_3^{(6)}=2\alpha \beta e^{\Phi} F_3^{(6)},
\ee
\be\label{W3H6}
(\alpha^2+\beta^2)W_3^{(6)}=-2\alpha \beta \star_{6} H_3^{(6)},
\ee
where the subscript in the Hodge star indicates that it is taken with respect to the metric in the extra 6-d space. Recall that all the fluxes (except for the components in the second term in the self-dual $\tilde{F}_5=(1+\star)F_5$) and the torsion have only internal components.
Combining (\ref{W3F6}) and (\ref{W3H6}) gives the relation between
$H_3^{(6)}$ and $F_3^{(6)}$,
\be\label{H6F6}
(\alpha^2-\beta^2)H_3^{(6)}=(\alpha^2+\beta^2) e^{\Phi}
\star_6 F_3^{(6)}.
\ee

\subsection{The $3\oplus\bar{3}$ sector}

This sector contains the equations for the running of the dilaton and the warp factor and all the fluxes have components in the $3\oplus\bar{3}$ representation.
The supersymmetry conditions in (\ref{QibWH})-(\ref{TibA}) together with the second set of constraints which can be read off with the change of variables  $\alpha \leftrightarrow \beta$,  $F^{(\bar{3})}_{1} \rightarrow -F^{(\bar{3})}_{1}$, $F^{(\bar{3})}_{3} \rightarrow -F^{(\bar{3})}_{3}$ and $F^{(\bar{3})}_{5} \rightarrow -F^{(\bar{3})}_{5}$ give the following constraints on the relations among the flux and the torsion components in the $\bar{3}$ representation, the running of the dilaton, the running of the warp factor, and the spinors parameters $\alpha$ and $\beta$,
\be\label{ththb1}
\partial_{\bar{k}}\alpha+\frac{1}{2}(\alpha(W^{(\bar{3})}_5-W^{(\bar{3})}_4)-i\beta H_3^{(\bar{3})})_{\bar{k}}=0,
\ee
\be\label{ththb2}
\partial_{\bar{k}}\beta+\frac{1}{2}(\beta(W^{(\bar{3})}_5-W^{(\bar{3})}_4)-i\alpha H_3^{(\bar{3})})_{\bar{k}}=0,
\ee
\be
\alpha\partial_{\bar{k}}A+\frac{i}{4}e^{\Phi}(\alpha F_1^{(\bar{3})}-2i\beta F_3^{(\bar{3})}-2\alpha F_5^{(\bar{3})})_{\bar{k}}=0,
\ee
\be
\beta\partial_{\bar{k}}A-\frac{i}{4}e^{\Phi}(\beta F_1^{(\bar{3})}-2i\alpha F_3^{(\bar{3})}-2\beta F_5^{(\bar{3})})_{\bar{k}}=0,
\ee
\be
(\alpha W^{(\bar{3}) }_4 -i \beta H_3^{(\bar 3)})_{\bar{k}}
- \frac{i}{2} e^{\Phi} (
\alpha F^{(\bar{3}) }_1
+2 \alpha F_5^{(\bar{3})})_{\bar{k}}=0,
\ee
\be
(\beta W^{(\bar{3}) }_4 -i \alpha H_3^{(\bar 3)})_{\bar{k}}
+ \frac{i}{2} e^{\Phi} (
\beta F^{(\bar{3}) }_1
+2 \beta F_5^{(\bar{3}) })_{\bar{k}}=0,
\ee
\be\label{ththba1}
\alpha \partial_{\bar{k}}(2A-\Phi+\ln \alpha)+\frac{1}{2}(\alpha(W^{(\bar{3})}_4+W^{(\bar{3})}_5)-i\beta H_3^{(\bar{3})})_{\bar{k}}=0,
\ee
\be\label{ththbf}
\beta \partial_{\bar{k}}(2A-\Phi+\ln \beta)+\frac{1}{2}(\beta(W^{(\bar{3})}_4+W^{(\bar{3})}_5)-i\alpha H_3^{(\bar{3})})_{\bar{k}}=0.
\ee
There are a total of eight constraints above for eight components of fluxes, torsion, and variations of the dilaton and the warp factor: $F_5^{(\bar{3})}$, $F_3^{(\bar{3})}$, $F_1^{(\bar{3})}$, $H_3^{(\bar{3})}$, $W_5^{(\bar{3})}$, $W_4^{(\bar{3})}$, $\partial_{\bar{k}}\Phi$ and $\partial_{\bar{k}}A$.  The objective is to find their relations to the spinors parameters $\alpha$ and $\beta$.
Solving the first two constraints (\ref{ththb1}) and (\ref{ththb2}) gives the same expression for $H_3^{(\bar{3})}$ as does solving the last two constrains (\ref{ththba1}) and (\ref{ththbf}). Therefore, the above  eight constrains actually give only seven independent complex equations.
The equations which solve the constraints (\ref{ththb1})-(\ref{ththbf}) are
\be
e^{\Phi}\Bigl(F_3^{(\bar{3})}+\frac{2i\alpha \beta}{\alpha^2+\beta^2}F_1^{(\bar{3})}\Bigr)=\frac{2\alpha \beta}{\alpha^2+\beta^2}(\bar{\partial}\ln \alpha -\bar{\partial}\ln \beta),\label{F33b}
\ee
\be
e^{\Phi}\Bigl(F_5^{(\bar{3})}+\frac{1}{2}F_1^{(\bar{3})}\Bigr)=-i(\bar{\partial}\ln \alpha -\bar{\partial}\ln \beta),
\ee
\be
H_3^{(\bar{3})}=\frac{2i\alpha \beta}{\alpha^2-\beta^2}(\bar{\partial}\ln \alpha -\bar{\partial}\ln \beta),\label{H33bg}
\ee
\be
\bar{\partial}A+\frac{i}{2}\frac{\alpha^2- \beta^2}{\alpha^2+\beta^2}e^{\Phi}F_1^{(\bar{3})}=\frac{\alpha^2- \beta^2}{2(\alpha^2+\beta^2)}(\bar{\partial}\ln \alpha -\bar{\partial}\ln\beta),\label{dbarA}
\ee
\be\label{edphi}
\bar{\partial}\Phi+i\frac{\alpha^2- \beta^2}{\alpha^2+\beta^2}e^{\Phi}F_1^{(\bar{3})}=-\frac{4\alpha^2 \beta^2}{\alpha^4-\beta^4}(\bar{\partial}\ln \alpha -\bar{\partial}\ln\beta),
\ee
\be
W_4^{(\bar{3})}=-\frac{\alpha^2+ \beta^2}{\alpha^2-\beta^2}(\bar{\partial}\ln \alpha -\bar{\partial}\ln \beta),\label{w4eom}
\ee
\be
W_5^{(\bar{3})}=-\frac{3\alpha^2 +\beta^2}{\alpha^2-\beta^2}\bar{\partial}\ln \alpha+\frac{\alpha^2 +3\beta^2}{\alpha^2-\beta^2}\bar{\partial}\ln \beta,\label{W5eom}
\ee
where the antiholomorphic index is now suppressed.
The above equations are general and allow to scan generic relations between the two Weyl spinors.

We can combine some of the equations to find relations between the different components. We can, for instance, express $F_1^{(\bar{3})}$ in terms of other fluxes using (\ref{F33b}) and (\ref{H33bg}),
\be
2\alpha\beta e^{\Phi}F_1^{(\bar{3})}=-({\alpha^2-\beta^2})H_3^{(\bar{3})}
+i({\alpha^2+\beta^2})e^{\Phi}F_3^{(\bar{3})}
\ee
which implies that a nonzero $F_1^{(\bar{3})}$ with nonzero $\alpha\beta$ requires that at least $H_3^{(\bar{3})}$ or $F_3^{(\bar{3})}$ be nonzero and, therefore, both 3-form fluxes cannot be primitive.
We can also relate $\bar{\partial}A$, $\bar{\partial}\Phi$ and $W_4^{(\bar{3})}$ by combining their expressions,
\be
2\bar{\partial}A-\bar{\partial}\Phi=-W_4^{(\bar{3})}
\ee
which, for instance, implies that a nonconstant warp factor with constant dilaton requires the torsion component $W_4^{\bar{3}}\ne0$. On the other hand, if $W_4^{\bar{3}}=0$, then we have $\bar{\partial}\Phi=2\bar{\partial}A$.

The warp factor is directly related to the magnitudes of the spinors parameters. The parametrization of $\eta_{+}^{i}$ in terms of $\eta_+$ given by (\ref{eta1eta2-def}) gives
\be
{\eta_{+}^1}^\dagger \eta_{+}^1+{\eta_{+}^2}^\dagger \eta_{+}^2=\frac{1}{2}(|\alpha|^2+|\beta|^2)\eta_+^\dagger \eta_+.\label{etapm-norm}
\ee
Normalizing the invariant spinor such that $\eta_+^\dagger \eta_+=1/2$ and recalling that
the 6-d spinors $\eta_{\pm}^1$ were defined in terms of $\eta_+$ with the warp factor which comes in the 4-d spinors absorbed in, (\ref{etapm-norm}) can be set to $e^{A}/4$. The warp factor is then related to $\alpha$ and $\beta$ as \cite{Grana:2004bg,Dall'Agata:2004dk,Frey:2004rn}
\be\label{Aalbe}
A=\ln(|\alpha|^2+|\beta|^2).
\ee

\section{Specific classes of supergravity flows}

We note that the equations given by (\ref{F33b})-(\ref{W5eom}) (except the one for $W_5^{(\bar{3})}$) are invariant under rotations of both $\alpha$ and $\beta$ by the same phase, $\alpha\to e^{i\delta}\alpha$ and $\beta\to e^{i\delta}\beta$. For $W_5^{(\bar{3})}$, this rotation leads to $W_5^{(\bar{3})}\to W_5^{(\bar{3})}-2i\bar{\partial}\delta$ and the shift in $W_5^{(\bar{3})}$ arises only when $\delta$ is not constant. We will take the phase of $\alpha$ to be a fixed value, and
what matters is then the magnitude of the phase between $\alpha$ and $\beta$ rather than the overall orientation. So, we can take
\be\alpha=|\alpha|,\qquad \beta=|\beta|e^{i\theta},\label{albeth}\ee where $\theta$ is real.
The equations for all sectors are then invariant and depend not on the overall phase but only on the phase between $\alpha$ and $\beta$.
We also have from (\ref{Aalbe}) that
$A=\ln(\alpha^2+\beta^2 e^{-i 2\theta})$ which with the equation for $\bar{\partial}A$ given by (\ref{dbarA}) gives
\be\label{dbedaldthg}
\bar{\partial}\ln \beta= \frac{\left(\frac{\alpha^2- \beta^2}{2(\alpha^2+\beta^2)}-\frac{2\alpha^2}{\alpha^2+\beta^2e^{-2i\theta}}\right)
\bar{\partial}\ln \alpha+ \frac{2i\beta^2e^{-2i\theta}}{\alpha^2+\beta^2e^{-2i\theta}}\bar{\partial}\theta-\frac{i}{2}\frac{\alpha^2- \beta^2}{\alpha^2+\beta^2}e^{\Phi}F_1^{(\bar{3})}}
{\frac{\alpha^2- \beta^2}{2(\alpha^2+\beta^2)}+\frac{2\beta^2e^{-2i\theta}}{\alpha^2+\beta^2e^{-2i\theta}}}.
\ee
With this, the equations given by (\ref{F33b})-(\ref{W5eom}) can be rewritten in terms of $\alpha$, $\beta$, $\bar{\partial}\ln \alpha$ and $\bar{\partial}\ln \theta$.
Since we need at least either one of $\alpha$ or $\beta$ to be nonzero in order to have nontrivial background with flux, we will take in our discussions that $\alpha\ne0$ while $\beta$ could vanish and write \be\frac{\beta}{\alpha}=\tan{(w/2)\,}e^{i\theta},\label{betaoalpha1}\ee where $w$ and $\theta$ are real.

Recall that $\alpha$ and $\beta$ define the relations between the spinors,
$\eta_+^1=\frac{1}{2}(\alpha+\beta) \eta_+$ and
$\eta_+^2=\frac{1}{2i}(\alpha-\beta) \eta_+$.  Thus ${\eta_{+}^2}/{\eta_{+}^1}=-i(\alpha-\beta)/(\alpha+\beta)$, and with (\ref{betaoalpha1}),
\be
\frac{\eta_{+}^2}{\eta_{+}^1}=-i\,\frac{\alpha-\beta}{\alpha+\beta}
=-i\frac{1-\tan{(w/2)}e^{i\theta}}{1+\tan{(w/2)}e^{i\theta}}
.\label{et2overet1}
\ee
By varying both $w$ and $\theta$, and using appropriate ansatz for the fluxes and the metric which accommodates the corresponding flow, one can scan the moduli space of type IIB backgrounds with $\mathcal{N}=1$ supersymmetry.
For generic values of $w$ and $\theta$, the ratio of the two spinors has neither constant magnitude nor constant phase, the backgrounds involve fluxes which are not imaginary self-dual, the 3-form fluxes are not primitive, and the dilaton and the axion fields are not constants.
Next we consider some specific classes of flows and discuss their features and relations.

\subsection{Flows with constant axion}

Let us consider the case in which the axion field $C_0$ is constant and set $F_1^{(\bar{3})}=0$. In this case, the equations in the $3\oplus \bar{3}$ sector, (\ref{F33b})-(\ref{W5eom}) together with (\ref{Aalbe}), become
\be\label{F35th0}
e^{\Phi}F_3^{(\bar{3})}=\frac{2\alpha \beta}{\alpha^2+\beta^2}(\bar{\partial}\ln \alpha -\bar{\partial}\ln \beta),
\quad
e^{\Phi}F_5^{(\bar{3})}=-i(\bar{\partial}\ln \alpha -\bar{\partial}\ln \beta),
\ee
\be
H_3^{(\bar{3})}=\frac{2i\alpha \beta}{\alpha^2-\beta^2}(\bar{\partial}\ln \alpha -\bar{\partial}\ln \beta),
\quad
W_4^{(\bar{3})}=-\frac{\alpha^2+ \beta^2}{\alpha^2-\beta^2}(\bar{\partial}\ln \alpha -\bar{\partial}\ln \beta),
\ee
\be\label{dAF10}
\bar{\partial}A=\frac{\alpha^2- \beta^2}{2(\alpha^2+\beta^2)}(\bar{\partial}\ln \alpha -\bar{\partial}\ln\beta), \quad A=\ln(|\alpha|^2+|\beta|^2),
\ee
\be\label{edphith0}
\bar{\partial}\Phi=-\frac{4\alpha^2 \beta^2}{\alpha^4-\beta^4}(\bar{\partial}\ln \alpha -\bar{\partial}\ln\beta),\quad
W_5^{(\bar{3})}=-\frac{3\alpha^2 +\beta^2}{\alpha^2-\beta^2}\bar{\partial}\ln \alpha+\frac{\alpha^2 +3\beta^2}{\alpha^2-\beta^2}\bar{\partial}\ln \beta.
\ee
Note that flows with constant axion but nonconstant dilaton have nonprimitive 3-form fluxes.
The equations in the $6\oplus \bar{6}$ sector do not directly involve $F_1$ and stay the same.

\subsubsection{Spinors with a phase difference of $\pi/2$ (GMPT case)}

Consider the specific case of a flow with constant axion and the relation between the two spinors such that $\theta=0$.
This case gives the equations obtained by Grana, Minasian, Petrini and Tomasiello (GMPT) in \cite{Grana:2004bg}. The argument used in \cite{Grana:2004bg} to obtain the equations was a gauge choice such that $\arg (\alpha)+\arg (\beta)=0$. As we see can see in the equations in \cite{Grana:2004bg}, given by (\ref{alpharbetar})-(\ref{edphirr}) below, the ratio of the expression for $W_5^{(\bar{3})}$ to the expressions for all the other components depends only on the phase between $\alpha$ and $\beta$.
Once the overall phase is fixed to a constant, we can write $\alpha$ and $\beta$ as given by (\ref{albeth}) without loss of generality. The gauge choice with (\ref{albeth}) is, then, equivalent to a phase of zero between $\alpha$ and $\beta$ (and $\alpha$ and $\beta$ being real-proportional).
Now we have from (\ref{et2overet1}) that ${\eta_{+}^2}/{\eta_{+}^1}=i {(\tan (w/2)-1)/(\tan (w/2)+1)}$ and the phase between the two spinors is a constant $\pi/2$.
We then have from (\ref{Aalbe}) that $A=\ln(\alpha^2+\beta^2)$ which with the expression for $\bar{\partial}A$ in (\ref{dAF10}), or directly reading off from (\ref{dbedaldthg}) with $\theta=0$ and $F_1^{(\bar{3})}=0$, gives
\be\label{alpharbetar}
\bar{\partial}\ln \beta=-\frac
{3\alpha^2+\beta^2}{\alpha^2+3\beta^2}\bar{\partial}\ln \alpha.
\ee
This in (\ref{F35th0})-(\ref{edphith0}) gives
\be
e^{\Phi}F_3^{(\bar{3})}=\frac{8\alpha\beta}
{\alpha^2+3\beta^2}\bar{\partial}\ln \alpha,
\quad
e^{\Phi}F_5^{(\bar{3})}=-\frac{4i(\alpha^2+\beta^2)}
{\alpha^2+3\beta^2}\bar{\partial}\ln \alpha,
\ee
\be
H_3^{(\bar{3})}=\frac{8i\alpha\beta(\alpha^2+\beta^2)}
{(\alpha^2+3\beta^2)(\alpha^2-\beta^2)}\bar{\partial}\ln \alpha,
\quad
W_4^{(\bar{3})}=-\frac{4(\alpha^2+\beta^2)^2}
{(\alpha^2+3\beta^2)(\alpha^2-\beta^2)}\bar{\partial}\ln \alpha,
\ee
\be\
\bar{\partial}A=\frac{2(\alpha^2-\beta^2)}
{\alpha^2+3\beta^2}\bar{\partial}\ln \alpha,
\quad A=\ln(\alpha^2+\beta^2),
\ee
\be\label{edphirr}
\bar{\partial}\Phi=-\frac{16\alpha^2 \beta^2}
{(\alpha^2+3\beta^2)(\alpha^2-\beta^2)}\bar{\partial}\ln \alpha, \quad W_5^{(\bar{3})}=-\frac{2(3\alpha^2+\beta^2)}
{\alpha^2-\beta^2}\bar{\partial}\ln \alpha.
\ee
These are the equations obtained in \cite{Grana:2004bg}, expressed in terms of $\alpha$, $\beta$ and $\bar{\partial}\ln \alpha$ here. Thus the equations written in \cite{Grana:2004bg} for the $3\oplus \bar{3}$ sector are a specific case of the general equations with a fixed phase of zero between $\alpha$ and $\beta$.

\subsubsection{Spinors with equal magnitude (BGMPZ case)} 

Consider the case in which the axion is constant and
a fixed phase of $\theta=\pi/2$ between $\alpha$ and $\beta$.  This gives the equations obtained by Butti, Grana, Minasian, Petrini, and Zaffaroni (BGMPZ) in \cite{Butti:2004pk}.
Note that, in this case, we have from (\ref{et2overet1}) that ${\eta_{+}^2}/{\eta_{+}^1}=-\sin w -i \cos w$. Therefore, the two spinors have equal magnitude. The phase between the spinors is $tan^{-1}(\cot w)$ and varies along the flow.
We can take $\alpha$ real and $\beta$ pure imaginary. In this case, (\ref{dbedaldthg}) reduces to
\be\label{alpharbetai}
\bar{\partial}\ln \beta=-\frac
{3\alpha^4+6\alpha^2\beta^2-\beta^4}{\alpha^4-6\alpha^2\beta^2-3\beta^4}\bar{\partial}\ln \alpha.
\ee
This in (\ref{F35th0})-(\ref{edphith0}) gives
\be
e^{\Phi}F_3^{(\bar{3})}=\frac{8\alpha\beta(\alpha^2-\beta^2)}
{\alpha^4-6\alpha^2\beta^2-3\beta^4}\bar{\partial}\ln \alpha,
\quad
e^{\Phi}F_5^{(\bar{3})}=-\frac{4i(\alpha^4-\beta^4)}
{\alpha^4-6\alpha^2\beta^2-3\beta^4}\bar{\partial}\ln \alpha,
\ee
\be
H_3^{(\bar{3})}=\frac{8i\alpha\beta(\alpha^2+\beta^2)}
{\alpha^4-6\alpha^2\beta^2-3\beta^4}\bar{\partial}\ln \alpha,
\quad
W_4^{(\bar{3})}=-\frac{4(\alpha^2+\beta^2)^2}
{\alpha^4-6\alpha^2\beta^2-3\beta^4}\bar{\partial}\ln \alpha,
\ee
\be
\bar{\partial}A=\frac{2(\alpha^2-\beta^2)^2}
{\alpha^4-6\alpha^2\beta^2-3\beta^4}\bar{\partial}\ln \alpha,\quad A=\ln(\alpha^2-\beta^2)
\ee
\be\label{edphiri}
\bar{\partial}\Phi=-\frac{16\alpha^2\beta^2}
{\alpha^4-6\alpha^2\beta^2-3\beta^4}\bar{\partial}\ln \alpha, \quad
W_5^{(\bar{3})}=-\frac{2(3\alpha^4+2\alpha^2\beta^2+3\beta^4)}
{\alpha^4-6\alpha^2\beta^2-3\beta^4}\bar{\partial}\ln \alpha.
\ee
These equations were written and used in \cite{Butti:2004pk} to study flows from the Klebanov-Strassler solution \cite{Klebanov:2000hb} toward the Maldacena-Nunez solution \cite{Maldacena:2000yy, Chamseddine:1997nm} using a metric and flux ansatz given in \cite{Papadopoulos:2000gj}.

\subsection{Conformally Calabi-Yau flows\label{sec-ccy}}

Conformally Calabi-Yau backgrounds are those for which the metric on the extra space can be written as $$ds_6^2 = e^{2B(y)}h_{mn}(y)dy^m dy^n,$$ where $h_{mn}(y)$ is a Calabi-Yau metric, and, therefore, one can make a conformal transformation to a Calabi-Yau metric.
In order for the metric on $Y$ to be conformally Calabi-Yau, we need $3W_4^{(\bar{3})}=2W_5^{(\bar{3})}$, see \cite{Gurrieri:2002wz} for instance. We see from (\ref{w4eom}) and (\ref{W5eom}) that this is the case if
\be
\bar{\partial}\ln \beta= -\frac{3\alpha^2-\beta^2}{\alpha^2-3\beta^2}\bar{\partial}\ln \alpha.\label{dbedal-ccy}
\ee
The remaining condition we need to make $Y$ conformally Calabi-Yau is the vanishing of the torsion class in the $6\oplus\bar{6}$ sector, $W_3^{(6)}=0$, which holds for $\beta=0$ and $\alpha\ne0$ (for nontrivial background with flux). Setting $\beta=0$ in (\ref{dbedal-ccy}), we have $\bar{\partial}\ln \beta=-3 \bar{\partial}\ln \alpha.$ On the other hand, using $\bar{\partial}\ln \beta$ given by (\ref{dbedaldthg}) with $\beta=0$,
\be\bar{\partial}\ln \beta=-3\bar{\partial}\ln \alpha-ie^{\Phi}F_1^{(\bar{3})}.\label{dlbedlalb01}\ee
Therefore, we need to have $F_1^{(\bar{3})}=0$.
The equations in the $3\oplus \bar{3}$ sector, (\ref{F33b})-(\ref{W5eom}), then reduce to
\be
W_4^{(\bar{3})}=-4\bar{\partial}\ln \alpha,\quad W_5^{(\bar{3})}=-6\bar{\partial}\ln \alpha,\quad
F_3^{(\bar{3})}=0,\quad H_3^{(\bar{3})}=0,\quad F_1^{(\bar{3})}=0\label{ccy-a}\ee\be\label{f5dphdaccy}
F_5^{(\bar{3})}=-4i \bar{\partial}\ln \alpha,\quad
\bar{\partial}\Phi=0,\quad  \bar{\partial}A=2\bar{\partial}\ln \alpha.
\ee
The equations in the $6\oplus \bar{6}$ sector given by (\ref{W3F6}) and (\ref{W3H6}) also reduce, in this case, to
\be
H_3^{(6)}= e^{\Phi} \star_6 F_3^{(6)},\quad W_3^{(6)}=0.\label{ccy-f}
\ee
Note that we see from (\ref{f5dphdaccy}) that the running of the warp factor and the 5-form flux in conformally Calabi-Yau flows are related as
\be
\bar{\partial}A=\frac{i}{2}e^{\Phi}F_5^{(\bar{3})}.\label{dAF5}
\ee

Therefore, conformally Calabi-Yau flows require that the 3-form fluxes be primitive, $F_3^{(\bar{3})}= H_3^{(\bar{3})}=0$, and have constant dilaton and constant axion.

Compactifications over Calabi-Yau threefolds give $\mathcal{N}=2$ supersymmetry in four dimensions. Nevertheless, let us see  in terms of the fluxes and the spinors parameters how flux backgrounds with $\mathcal{N}=1$ supersymmetry are necessarily non-Calabi-Yau. In order to have a Calabi-Yau flow, the fundamental 2-form and the holomorphic 3-form need to be closed and consequently all the torsion classes need to vanish.
This requires setting $\bar{\partial}\ln \alpha=0$ in the equations for the conformally Calabi-Yau flows above. But, then, the 5-form flux vanishes. With (\ref{ccy-f}) and because we have from the bosonic type IIB supergravity equations that $dF_5=F_3\wedge H_3$, the 3-form fluxes vanish too (i.e., the components of the 3-form fluxes in the $6\oplus \bar{6}$ representation vanish in addition to the ones in the $3\oplus \bar{3}$). This corresponds to trivial background with a stationary point in the $(\alpha,\beta)$ spinors parameters space, all fluxes vanishing, and the dilaton and the warp factor being constants. In other words, backgrounds on such stationary points in the spinors parameters space reduce to Calabi-Yau.

\subsection{Flows with constant dilaton}

We see from the equation for the running of the dilaton given by (\ref{edphi}) that  the dilaton is a constant, $\bar{\partial}\Phi=0$, for flows such that
\be
4\alpha^2\beta^2\bar{\partial}\ln \beta= 4\alpha^2\beta^2 \bar{\partial}\ln \alpha + ie^{\Phi}(\alpha^2-\beta^2)^2 F_1^{(\bar{3})}
\ee
which with (\ref{F33b})-(\ref{W5eom}) gives
\be\label{F3W4W5b}
{2\alpha\beta}F_3^{(\bar{3})}+{i(\alpha^2+\beta^2) F_1^{(\bar{3})}}
=0,
\ee \be
{4\alpha^2\beta^2}F_5^{(\bar{3})}
+{(\alpha^4+\beta^4)F_1^{(\bar{3})}}=0
,\quad{2\alpha\beta} H_3^{(\bar{3})}-{e^{\Phi}(\alpha^2-\beta^2) F_1^{(\bar{3})}}
=0,
\ee
\be {8\alpha^2\beta^2}\bar{\partial}A+{ie^{\Phi}(\alpha^4-\beta^4) F_1^{(\bar{3})}}
=0,\quad \quad A=\ln(|\alpha|^2+|\beta|^2)\ee
\be
{4\alpha^2\beta^2}W_4^{(\bar{3})}-{ie^{\Phi}(\alpha^4-\beta^4) F_1^{(\bar{3})}}
=0, \ee\be\label{edAirrW5dPh}
{4\alpha^2\beta^2}W_5^{(\bar{3})}-{ie^{\Phi}(\alpha^4+2\alpha^2\beta^2-3\beta^4) F_1^{(\bar{3})}}
=-{8\alpha^2\beta^2}\bar{\partial}\ln \alpha.
\ee
Note that flows with constant dilaton but nonconstant axion have nonprimitive 3-form fluxes.

This class of flows with constant dilaton contains the specific case  of  $\beta=0$ or $\alpha=0$ with $F_1^{(\bar{3})}=0$ discussed in \cite{Butti:2004pk}.
For the specific case of  $\beta=0$ and $F_1^{(\bar{3})}=0$, we use (\ref{dbedaldthg}) in (\ref{F33b})-(\ref{Aalbe}), with $\Phi=0$, and the equations in the $3\oplus\bar{3}$ sector reduce to the ones given by (\ref{ccy-a}) and (\ref{f5dphdaccy}).
Thus flows with constant dilaton and constant axion require that the 3-form fluxes in the $3\oplus \bar{3}$ representation vanish (the 3-form fluxes be primitive), and the background be conformally Calabi-Yau. We also note that backgrounds with only $F_5$ flux fall in this class with all components of the 3-form fluxes set to zero and the running of the warp factor related to the 5-form flux as given by (\ref{dAF5}).

However, we note from the equations above that
there is a much larger region of parameters space with nonzero $F_1^{(\bar{3})}$, nonzero components of the 3-form fluxes in the $3\oplus \bar{3}$ representation
and nonzero $\alpha$ and $\beta$ which gives flows with constant dilaton.

\subsection{Flows with imaginary self-dual 3-form flux\label{sec-isd}}

Now we like to see features of flows with imaginary self-dual combination of the 3-form fluxes. This constraint is often assumed in IIB flux compactifications, for instance in \cite{Giddings:2001yu}. Consider the combination of the 3-form fluxes
\be
G_3=F_3-ie^{-\Phi} H_3.
\ee
The 3-form combination
$G_3$ is imaginary self-dual if $\star_6 G_3=i G_3$ which implies
\be\star_6 H_3=-e^{\Phi} F_3.\ee
Noting from (\ref{H3-1a}) and (\ref{F3-1a}) that $H_3\supset H_3^{(\bar{3})}\wedge J$ and  $F_3\supset F_3^{(\bar{3})}\wedge J$, the imaginary self-duality constraint implies that for the components in the $\bar{3}$ representation, \be
\star_6 (H_3^{(\bar{3})}\wedge J)=- e^{\Phi} F_3^{(\bar{3})}\wedge J.\label{isd6}\ee
The imaginary self-duality constraint for the components in the $6$ representation, $\star_6 H_3^{(6)}=-e^{\Phi} F_3^{(6)}$,
together with (\ref{H6F6})  gives
\be
\frac{\alpha^2+\beta^2}{\alpha^2-\beta^2}=1.\label{isdalbe}
\ee
Now using the equation for the components in the $\bar{3}$ representation given by (\ref{F33b}) and (\ref{H33bg}),
\be
{H_3^{(\bar{3})}}=i{e^{\Phi}}\frac{\alpha^2+\beta^2}{\alpha^2-\beta^2}{F_3^{(\bar{3})}}
=ie^{\Phi}{F_3^{(\bar{3})}},\label{isd3b}
\ee
where we have used (\ref{isdalbe}) in the last step.
Using (\ref{isd3b}) in (\ref{isd6}),
\be
\star_6 (H_3^{(\bar{3})}\wedge J)=i H_3^{(\bar{3})}\wedge J.\label{isd3b3}\ee

On the other hand, taking the Hodge star directly,
\be\star_6 ((H_3^{(\bar{3})})\wedge J)=-i H_3^{(\bar{3})}\wedge J.\label{isd3b3c}\ee
In taking the Hodge star, we used the fact that the components of the total fluxes, $H_3$ and $F_3$, are real.

Thus it follows from (\ref{isd3b3}) and (\ref{isd3b3c}) that $H_3^{(\bar{3})}=-H_3^{(\bar{3})}$. Similarly, repeating the same proof for $F_3$, we need $F_3^{(\bar{3})}=-F_3^{(\bar{3})}$. Therefore, the imaginary self-duality constraint implies that the components of the 3-form fluxes in the $3\oplus \bar{3}$ representation must vanish.
For both $H_3^{(\bar{3})}$ and $F_3^{(\bar{3})}$ to vanish, we see from (\ref{F33b}), (\ref{H33bg}) and the equations in the $6\oplus\bar{6}$ that we need $\alpha \beta=0$.
But in order to have nontrivial flow with flux, we need at least either one of $\alpha$ or $\beta$ to be nonzero. We can take $\beta=0$ and $\alpha\ne0$.
Putting $\beta=0$ in (\ref{dbedaldthg}), we find
\be
\bar{\partial}\ln \alpha-\bar{\partial}\ln \beta=4 \bar{\partial}\ln \alpha+i e^{\Phi}F_1^{(\bar{3})}.\label{dlaldbeb0}
\ee
Using (\ref{dlaldbeb0}) in the equations in the $3\oplus\bar{3}$ sector (\ref{F33b})-(\ref{W5eom}),
\be
e^{\Phi}F_3^{(\bar{3})}=H_3^{(\bar{3})}=0,
\quad
e^{\Phi}F_5^{(\bar{3})}=-4i\bar{\partial}\ln \alpha+\frac{1}{2}e^{\Phi}F_1^{(\bar{3})},
\ee
\be
\quad
W_4^{(\bar{3})}=-4\bar{\partial}\ln \alpha-i e^{\Phi}F_1^{(\bar{3})},\quad W_5^{(\bar{3})}=-6\bar{\partial}\ln \alpha
\ee
\be
\bar{\partial}\Phi=-i e^{\Phi}F_1^{(\bar{3})},\quad \bar{\partial}A=2\bar{\partial}\ln \alpha.
\ee
In addition, the equations in the $6\oplus\bar{6}$ sector reduce to $H_3^{(6)}= e^{\Phi} \star_6 F_3^{(6)}$ and $W_3^{(6)}=0$.
For constant axion, we have $3W_4^{(\bar{3})}=2W_5^{(\bar{3})}$ in addition to  $W_3^{(6)}=0$. Therefore, flows with imaginary self-dual $G_3$ and constant axion are conformally Calabi-Yau. Moreover, for flows with imaginary self-dual $G_3$ and constant axion,  the dilaton is constant. Note that the dilaton-axion coupling $ie^{-\Phi}+iC_0$ is constant for flows with imaginary self-dual $G_3$. Similar statements can be made for imaginary anti-self-dual fluxes ($\star_6 G_3=-i G_3$) with corresponding dilaton-axion coupling coefficient $-ie^{-\Phi}+C_0$.

The Klebanov-Strassler solution is an example with constant axion, constant dilaton, and imaginary self-dual 3-form flux on a conformally Calabi-Yau background.

\subsection{Flows with spinors of fixed ratio}

The supersymmetry transformations simplify and the equations are readily found for the special cases of $\alpha=0$, $\beta=0$, $\beta=\pm \alpha$ or $\beta=\pm i\alpha$. Now we like to consider the more general case of $\beta/\alpha=constant$.
Let us write
\be\frac{\beta}{\alpha}=c,\label{bealc}\ee
where $c$ is a complex constant. The relations between the spinors given by (\ref{eta1eta2-def}) then become
\be \eta_+^1=\frac{1}{2}\alpha(1+c) \eta_+,\qquad
\eta_+^2=\frac{1}{2i}\alpha(1-c) \eta_+,\qquad \frac{\eta_+^2}{\eta_+^1}=-i\frac{1-c}{1+c}.\ee
We can find $\bar{\partial}\ln \alpha-\bar{\partial}\ln \beta$ using (\ref{bealc}) and the expressions for $\bar{\partial}A$ and $A$ given by (\ref{dbarA}) and (\ref{Aalbe}), or read it off from (\ref{dbedaldthg}),
\be
\bar{\partial}\ln \beta=\bar{\partial}\ln \alpha-\frac{2\bar{\partial}\ln \alpha+\frac{i}{2}\frac{1-c^2}{1+c^2}e^{\Phi}F_1^{(\bar{3})}}{\frac{i}{2}\frac{1-c^2}{1+c^2}
+\frac{2|c|^2}{1+|c|^2}}.\label{dlbdlac}
\ee
The equations in the $3\oplus \bar{3}$ sector then follow from (\ref{bealc}) and  (\ref{dlbdlac}) in (\ref{F33b})-(\ref{W5eom}) and (\ref{Aalbe}). The equations in the $6\oplus \bar{6}$ sector also follow from (\ref{bealc}) in (\ref{W3F6}) and (\ref{W3H6}).

Let us consider the more specific case of nonzero $c$ with both $\alpha$ and $\beta$ nonzero. In this case, we have $\bar{\partial}\ln \beta-\bar{\partial}\ln \alpha=0$ and the equations in the $3\oplus \bar{3}$ sector reduce to
\be
F_3^{(\bar{3})}=-\frac{2i\alpha \beta}{\alpha^2+\beta^2}F_1^{(\bar{3})},
\quad
F_5^{(\bar{3})}=-\frac{1}{2}F_1^{(\bar{3})},\quad H_3^{(\bar{3})}=0,\label{dbda0F3}
\ee
\be
\bar{\partial}A=-\frac{i}{2}\frac{\alpha^2- \beta^2}{\alpha^2+\beta^2}e^{\Phi}F_1^{(\bar{3})},
\quad A=\ln(|\alpha|^2+|\beta|^2),\label{dbda0A}
\ee
\be
\bar{\partial}\Phi=-i\frac{\alpha^2- \beta^2}{\alpha^2+\beta^2}e^{\Phi}F_1^{(\bar{3})},\quad
W_4^{(\bar{3})}=0,
\quad
W_5^{(\bar{3})}=-2\,\bar{\partial}\ln \alpha.\label{dbda0W5}
\ee
Moreover, it follows from (\ref{dlbdlac}) and $\bar{\partial}\ln \beta-\bar{\partial}\ln \alpha=0$ that
\be
e^{\Phi}F_1^{(\bar{3})}=4i \frac{1+c^2}{1-c^2} \bar{\partial}\ln \alpha.\label{F3-1c}
\ee
With (\ref{F3-1c}), (\ref{dbda0F3})-(\ref{dbda0W5}) reduce to
\be
e^{\Phi}F_3^{(\bar{3})}=-\frac{8c}{c^2-1}\bar{\partial}\ln \alpha,
\quad
e^{\Phi}F_5^{(\bar{3})}=\frac{2i(c^2+1)}{c^2-1}\bar{\partial}\ln \alpha,
\ee
\be
\bar{\partial}A=2\bar{\partial}\ln \alpha,
\quad
\bar{\partial}\Phi=4\bar{\partial}\ln \alpha,
\quad
W_5^{(\bar{3})}=-2\,\bar{\partial}\ln \alpha,
\ee
\be
\quad H_3^{(\bar{3})}=0,\quad W_4^{(\bar{3})}=0.\ee
The equations in the $6\oplus \bar{6}$ sector given by (\ref{W3F6}) and (\ref{W3H6}) also reduce to
\be\label{W3F6H6c}
(1-c^2)W_3^{(6)}=2c e^{\Phi} F_3^{(6)},
\qquad
(1+c^2)W_3^{(6)}=-2ic \star_6 H_3^{(6)}.
\ee
Note that this specific class of flows involves nonprimitive $F_3$ flux while the $H_3$ flux is primitive. Moreover, the flows of
the fields in the $\bar{3}$ sector are driven by the $W_5^{(\bar{3})}$ component of the torsion while $W_4^{(\bar{3})}=0$. For the 3-form fluxes in the $6$ representation, $H_3^{(6)}\ne e^{\Phi} \star_6  F_3^{(6)}$ and the difference is driven by the $W_3^{(6)}$ component of the torsion. Thus the 3-form fluxes do not give imaginary self-dual combination, the background geometry is not conformally Calabi-Yau, and the dilaton is not constant.
The Maldacena-Nunez solution \cite{Maldacena:2000yy} is an example in this class with $c=i$. In this case, $F_1^{(\bar{3})}=0$ and $F_5^{(\bar{3})}=0$ while $F_3^{(\bar{3})}$, the dilaton and the warp factor run and are driven by nonzero $W_5^{(\bar{3})}$ torsion. Moreover, $H_3^{(6)}=0$ while $F_3^{(6)}$ is balanced by $W_3^{(6)}$.

\section{Conclusions and discussions}

We have presented the general and explicit equations which solve the supersymmetry transformations with two arbitrary complex-proportional Weyl spinors on type IIB backgrounds with $\mathcal{N}=1$ supersymmetry and $SU(3)$ structures. These equations allow to study generic flows systematically with any of the fluxes turned on.

The equations can be used to read off and prove some features of  $\mathcal{N}=1$ supersymmetric type IIB backgrounds with $SU(3)$ structures. For instance,
\begin{itemize}
\item The flux and the torsion components in the singlet representation (the ones in the $(0,3)$ and $(3,0)$ forms) cannot balance each other and the singlet components of the 3-form fluxes must vanish identically.
\item Flows with constant axion, constant dilaton and nonconstant warp factor are conformally Calabi-Yau.
\item Conformally Calabi-Yau flows have imaginary self-dual 3-form flux.
\item Flows with imaginary self-dual 3-form flux have primitive 3-form fluxes.
\item Flows with imaginary self-dual 3-form flux have constant dilaton-axion coupling coefficient $\tau=ie^{-\Phi}+C_0$.
\end{itemize}
Moreover, other statements could be deduced from the above. For instance,
flows with imaginary self-dual 3-form flux and constant axion have constant dilaton.
Flows with imaginary self-dual 3-form flux and constant dilaton have constant axion.
 Flows with constant axion and nonconstant dilaton have nonprimitive 3-form flux.
Flows with constant dilaton and nonconstant axion have nonprimitive 3-form flux.
Conformally Calabi-Yau flows with constant axion have constant dilaton.
\comment{Moreover, other statements could be deduced from the above. For instance,
\begin{itemize}
\item Flows with imaginary self-dual 3-form flux and constant axion have constant dilaton.
\item Flows with imaginary self-dual 3-form flux and constant dilaton have constant axion.
\item Flows with constant axion and nonconstant dilaton have nonprimitive 3-form flux.
\item Flows with constant dilaton and nonconstant axion have nonprimitive 3-form flux.
\item Conformally Calabi-Yau flows with constant axion have constant dilaton.
\end{itemize}}
Some of these statements are familiar. Here, all the above relations follow and are proved using the equations. Some combinations of features listed here are often assumed in simplifying and analyzing IIB backgrounds. The relations could be helpful in distinguishing independent assumptions on backgrounds with $\mathcal{N}=1$ supersymmetry.

The richness of the flows which could be systematically studied using the equations is worth emphasizing. A background geometry may have different nonzero flux components along different directions and flows from one class to another are possible. A starting point in looking for explicit solutions may be picking up a particular direction in the spinors parameters space and writing down a metric and flux ansatz which accommodates the corresponding flow.
Particularly interesting could be gravity theories which are dual to gauge theories with physically interesting renormalization group flows such as pure confining gauge theories and those with flows suitable for inflationary cosmology.

For instance, the equations for the specific case in which the two spinors have equal magnitude ($\theta=\pi/2$) were used in \cite{Butti:2004pk} to study a flow from the Klebanov-Strassler solution \cite{Klebanov:2000hb} toward the Maldacena-Nunez solution \cite{Maldacena:2000yy} using a metric and flux ansatz given in \cite{Papadopoulos:2000gj}. However, it was learned that the flow hits a singularity where flux and torsion components blow up as one gets closer to a point where the Maldacena-Nunez solution is located in the $(\alpha,\, \beta)$ parameters space. This indicates to us that the flow with a fixed phase of $\theta=\pi/2$ and the ansatz in \cite{Papadopoulos:2000gj} is not suitable. In fact, we can see that the relation between $\bar{\partial}\ln \beta$ and $\bar{\partial}\ln \alpha$ given by (\ref{alpharbetai}) (and thus the relation between the two spinors) hits a singularity on this direction.
This can be avoided with flows along different directions using the general equations we have presented here with appropriate flux and metric ansatz.

The equations should also be useful to study supergravity flows with corrections to the anomalous mass dimension in the Klebanov-Strassler solution as the flow with the corrections included does not occur along a fixed phase between the spinors if the axion is constant or requires turning on the $F_1$ flux \cite{Hailu:2006uj}. This could be interesting because the corrections lead to distinct features in the warp factor and could be used to construct cosmological models which might possibly allow to probe stringy signatures from the early universe. The corrections make the dilaton run, and therefore, the background involves 3-form fluxes which do not form imaginary self-dual combination if the axion is constant.

In addition, the equations may be used to check if background solutions obtained through other means preserve $\mathcal{N}=1$ supersymmetry.

The backgrounds we have studied preserve $\mathcal{N}=1$ supersymmetry. Although supersymmetry can be easily broken by imbalance among the components of the fluxes, the torsion, the dilaton and the warp factor in any one of the equations, the $SU(3)$ singlet components of the 3-form fluxes seem to be natural candidates for breaking supersymmetry by flux. These components cannot be balanced by torsion as we have seen in the equations in the $1\oplus 1$ sector. Consequently, $F_3^{(1)}\ne 0$ or
$H_3^{(1)}\ne 0$ automatically breaks supersymmetry.
Another possibility for breaking supersymmetry is to turn on both imaginary self-dual and imaginary anti-self-dual fluxes which preserve different $\mathcal{N}=1$ subalgebras of the parent $\mathcal{N}=2$ type IIB theory compactified on Calabi-Yau manifold. A possibility for achieving metastable geometric configurations by wrapping $D5$- and anti-$D5$-branes on 2-spheres inside Calabi-Yau manifolds (which turn on imaginary self-dual and imaginary anti-self-dual fluxes respectively after geometric transition) was discussed in \cite{Aganagic:2006ex}.
The equations we have here could  be useful in searching for stable flux vacua with supersymmetry softly broken.
It could be interesting to investigate how the gravitino mass varies in warped compactifications with supersymmetry softly broken, as in the study in \cite{Burgess:2006mn} for instance, since the mass of the gravitino is an important parameter related to the Hubble constant when  applying supergravity backgrounds to construct cosmological models, see \cite{Kallosh:2007wm} for instance.

Finally, we would like to reemphasize that the equations should be useful in systematically searching for supergravity flows and vacua suitable for physical applications such as early universe cosmology, supersymmetry breaking, and gravity approaches to QCD, since only a very tiny subset of backgrounds has thus far been closely explored.

\section*{Acknowledgements}

We are very grateful to Henry Tye for insightful discussions. We thank Mariana Grana and  Alessandro Tomasiello for comments on the manuscript.
This research is supported in part by the National Science Foundation under grant number
NSF-PHY/03-55005.

\newpage

\appendix

\section{Summary of the general equations\label{appA}}

Here we summarize the general equations for type IIB backgrounds which preserve $\mathcal{N}=1$ supersymmetry on generalized Calabi-Yau with $SU(3)$ structures with all $F_1$, $F_3$, $F_5$ and $H_3$ fluxes turned on and arbitrary relation between the two complex-proportional Weyl spinors. The torsion components come in the variations of the fundamental 2-form, $dJ=-\frac{3}{2}\mathrm{Im}(W_1^{(1)} \bar{\Omega})+(W_4^{(3)}+W_4^{(\bar{3})})\wedge J+(W_3^{(6)}+W_3^{(\bar{6})})$ and the holomorphic 3-form,
$d\Omega=W_1^{(1)} J^2+W_2^{(8)}\wedge J+W_5^{(\bar{3})}\wedge \Omega$. The components of the 3-form fluxes come in $H_3=-\frac{3}{2}\mathrm{Im}(H_3^{(1)} \bar{\Omega})+(H_3^{(3)}+H_3^{(\bar{3})})\wedge J+(H_3^{(6)}+H_3^{(\bar{6})})$ and
$F_3=-\frac{3}{2}\mathrm{Im}(F_3^{(1)} \bar{\Omega})+(F_3^{(3)}+F_3^{(\bar{3})})\wedge J+(F_3^{(6)}+H_3^{(\bar{6})})$.
The self-dual 5-form flux is written as $\tilde{F}_5=(1+\star)F_5$ with
${F_5}=(F_5^{(3)}+F_5^{(\bar{3})})\wedge J\wedge J$.
The 1-form flux is decomposed as $F_1=F_1^{(\bar{3})}+F_1^{(3)}$. The superscripts denote the $SU(3)$ representations. The metric on the ten dimensional spacetime is written as $ds_{10}^2=e^{2A{(y)}}\eta_{\mu\nu}dx^\mu dx^\nu +ds_6^2(y)$ and the metric on the extra space is expressed in terms of complex 1-forms with holomorphic/antiholomorphic indices as $ds_6^2=\delta_{i\bar{j}}Z^{i}\bar{Z}^{\bar{j}}$.  The parameters $\alpha$ and $\beta$ are complex functions of the coordinates on the extra space and come in the relations between the two complex-proportional Weyl spinors $\eta_+^{1,2}$ and the $SU(3)$ invariant spinor $\eta_+$ on the generalized Calabi-Yau, $\eta_+^1=\frac{1}{2}(\alpha+\beta) \eta_+$ and
$\eta_+^2=\frac{1}{2i}(\alpha-\beta) \eta_+$. The equations which describe the balance among the components of the fluxes, the dilaton, the warp factor and the torsion in terms of the complex parameters $\alpha$ and $\beta$ such that $\mathcal{N}=1$ supersymmetry is preserved are summarized below.

\begin{center}
\begin{tabular}{|c|}
\hline
$\mathbf{1\oplus1}$ \textbf{and} $\mathbf{8\oplus8}$ \textbf{sectors}\tabularnewline
\hline
\hline
\T
$W^{(1)}_1=0,\quad W^{(8)}_2=0,\quad F^{(1)}_3=0,\quad H^{(1)}_3=0
$
\B
\tabularnewline
\hline
\end{tabular}
\end{center}

\begin{center}
\begin{tabular}{|c|}
\hline
$\mathbf{6\oplus\bar{6}}$ \textbf{sector}\tabularnewline
\hline
\hline
\T
$(\alpha^2-\beta^2)W_3^{(6)}=2\alpha \beta e^{\Phi} F_3^{(6)}$
\B
\tabularnewline
\hline
\T
$(\alpha^2+\beta^2)W_3^{(6)}=-2\alpha \beta \star_{6} H_3^{(6)}$
\B
\tabularnewline
\hline
\end{tabular}
\end{center}

\begin{center}
\begin{tabular}{|c|}
\hline
$\mathbf{3\oplus\bar{3}}$ \textbf{sector}\tabularnewline
\hline
\hline
\T
$e^{\Phi}\Bigl(F_3^{(\bar{3})}+\frac{2i\alpha \beta}{\alpha^2+\beta^2}F_1^{(\bar{3})}\Bigr)=\frac{2\alpha \beta}{\alpha^2+\beta^2}(\bar{\partial}\ln \alpha -\bar{\partial}\ln \beta)$\B\tabularnewline
\hline\T
$e^{\Phi}\Bigl(F_5^{(\bar{3})}+\frac{1}{2}F_1^{(\bar{3})}\Bigr)=-i(\bar{\partial}\ln \alpha -\bar{\partial}\ln \beta)$\B\tabularnewline
\hline\T
$H_3^{(\bar{3})}=\frac{2i\alpha \beta}{\alpha^2-\beta^2}(\bar{\partial}\ln \alpha -\bar{\partial}\ln \beta)\B$\tabularnewline
\hline
\T
$\bar{\partial}\Phi+i\frac{\alpha^2- \beta^2}{\alpha^2+\beta^2}e^{\Phi}F_1^{(\bar{3})}=-\frac{4\alpha^2 \beta^2}{\alpha^4-\beta^4}(\bar{\partial}\ln \alpha -\bar{\partial}\ln\beta)$\B\tabularnewline
\hline\T
$\bar{\partial}A+\frac{i}{2}\frac{\alpha^2- \beta^2}{\alpha^2+\beta^2}e^{\Phi}F_1^{(\bar{3})}=\frac{\alpha^2- \beta^2}{2(\alpha^2+\beta^2)}(\bar{\partial}\ln \alpha -\bar{\partial}\ln\beta),
\quad
A=\ln(|\alpha|^2+|\beta|^2)$\B\tabularnewline
\hline\T
$W_4^{(\bar{3})}=-\frac{\alpha^2+ \beta^2}{\alpha^2-\beta^2}(\bar{\partial}\ln \alpha -\bar{\partial}\ln \beta)$\B\tabularnewline
\hline\T
$W_5^{(\bar{3})}=-\frac{3\alpha^2 +\beta^2}{\alpha^2-\beta^2}\bar{\partial}\ln \alpha+\frac{\alpha^2 +3\beta^2}{\alpha^2-\beta^2}\bar{\partial}\ln \beta$\B\tabularnewline
\hline
\end{tabular}
\end{center}

\bibliographystyle{JHEP}


\begin{thebibliography}{10}

\bibitem{Grana:2004bg}
M.~Grana, R.~Minasian, M.~Petrini, and A.~Tomasiello, {\it Supersymmetric
  backgrounds from generalized calabi-yau manifolds},  {\em JHEP} {\bf 08}
  (2004) 046, [\href{http://xxx.lanl.gov/abs/hep-th/0406137}{{\tt
  hep-th/0406137}}].

\bibitem{Butti:2004pk}
A.~Butti, M.~Grana, R.~Minasian, M.~Petrini, and A.~Zaffaroni, {\it The
  baryonic branch of klebanov-strassler solution: A supersymmetric family of
  su(3) structure backgrounds},  {\em JHEP} {\bf 03} (2005) 069,
  [\href{http://xxx.lanl.gov/abs/hep-th/0412187}{{\tt hep-th/0412187}}].

\bibitem{Maldacena:1998re}
J.~Maldacena, {\it The large ${N}$ limit of superconformal field theories and
  supergravity},  {\em Adv. Theor. Math. Phys.} {\bf 2} (1998) 231--252,
  [\href{http://xxx.lanl.gov/abs/hep-th/9711200}{{\tt hep-th/9711200}}].

\bibitem{Gubser:1998bc}
S.~S. Gubser, I.~R. Klebanov, and A.~M. Polyakov, {\it Gauge theory correlators
  from non-critical string theory},  {\em Phys. Lett.} {\bf B428} (1998)
  105--114, [\href{http://xxx.lanl.gov/abs/hep-th/9802109}{{\tt
  hep-th/9802109}}].

\bibitem{Witten:1998qj}
E.~Witten, {\it Anti-de {S}itter space and holography},  {\em Adv. Theor. Math.
  Phys.} {\bf 2} (1998) 253--291,
  [\href{http://xxx.lanl.gov/abs/hep-th/9802150}{{\tt hep-th/9802150}}].

\bibitem{Giddings:2001yu}
S.~B. Giddings, S.~Kachru, and J.~Polchinski, {\it Hierarchies from fluxes in
  string compactifications},  {\em Phys. Rev.} {\bf D66} (2002) 106006,
  [\href{http://xxx.lanl.gov/abs/hep-th/0105097}{{\tt hep-th/0105097}}].

\bibitem{Kachru:2003aw}
S.~Kachru, R.~Kallosh, A.~Linde, and S.~P. Trivedi, {\it De sitter vacua in
  string theory},  {\em Phys. Rev.} {\bf D68} (2003) 046005,
  [\href{http://xxx.lanl.gov/abs/hep-th/0301240}{{\tt hep-th/0301240}}].

\bibitem{Dvali:1998pa}
G.~R. Dvali and S.-H.~H. Tye, {\it Brane inflation},  {\em Phys. Lett.} {\bf
  B450} (1999) 72, [\href{http://xxx.lanl.gov/abs/hep-ph/9812483}{{\tt
  hep-ph/9812483}}].

\bibitem{Kachru:2003sx}
S.~Kachru, R.~Kallosh, A.~Linde, J.~Maldacena, L.~McAllister, and S.~P.
  Trivedi, {\it Towards inflation in string theory},  {\em JCAP} {\bf 0310}
  (2003) 013, [\href{http://xxx.lanl.gov/abs/hep-th/0308055}{{\tt
  hep-th/0308055}}].

\bibitem{Tye:2006uv}
S.~H.~H. Tye, {\it Brane inflation: String theory viewed from the cosmos},
  \href{http://xxx.lanl.gov/abs/hep-th/0610221}{{\tt hep-th/0610221}}.

\bibitem{Vafa:2000wi}
C.~Vafa, {\it Superstrings and topological strings at large n},  {\em J. Math.
  Phys.} {\bf 42} (2001) 2798--2817,
  [\href{http://xxx.lanl.gov/abs/hep-th/0008142}{{\tt hep-th/0008142}}].

\bibitem{Gurrieri:2002wz}
S.~Gurrieri, J.~Louis, A.~Micu, and D.~Waldram, {\it Mirror symmetry in
  generalized calabi-yau compactifications},  {\em Nucl. Phys.} {\bf B654}
  (2003) 61--113, [\href{http://xxx.lanl.gov/abs/hep-th/0211102}{{\tt
  hep-th/0211102}}].

\bibitem{Strominger:1986uh}
A.~Strominger, {\it Superstrings with torsion},  {\em Nucl. Phys.} {\bf B274}
  (1986) 253.

\bibitem{deWit:1986xg}
B.~de~Wit, D.~J. Smit, and N.~D. Hari~Dass, {\it Residual supersymmetry of
  compactified d=10 supergravity},  {\em Nucl. Phys.} {\bf B283} (1987) 165.

\bibitem{Gauntlett:2002sc}
J.~P. Gauntlett, D.~Martelli, S.~Pakis, and D.~Waldram, {\it G-structures and
  wrapped ns5-branes},  {\em Commun. Math. Phys.} {\bf 247} (2004) 421--445,
  [\href{http://xxx.lanl.gov/abs/hep-th/0205050}{{\tt hep-th/0205050}}].

\bibitem{LopesCardoso:2002hd}
G.~Lopes~Cardoso {\em et.~al.}, {\it Non-kaehler string backgrounds and their
  five torsion classes},  {\em Nucl. Phys.} {\bf B652} (2003) 5--34,
  [\href{http://xxx.lanl.gov/abs/hep-th/0211118}{{\tt hep-th/0211118}}].

\bibitem{Gauntlett:2003cy}
J.~P. Gauntlett, D.~Martelli, and D.~Waldram, {\it Superstrings with intrinsic
  torsion},  {\em Phys. Rev.} {\bf D69} (2004) 086002,
  [\href{http://xxx.lanl.gov/abs/hep-th/0302158}{{\tt hep-th/0302158}}].

\bibitem{Grana:2005jc}
M.~Grana, {\it Flux compactifications in string theory: A comprehensive
  review},  {\em Phys. Rept.} {\bf 423} (2006) 91--158,
  [\href{http://xxx.lanl.gov/abs/hep-th/0509003}{{\tt hep-th/0509003}}].

\bibitem{Grana:2006kf}
M.~Grana, R.~Minasian, M.~Petrini, and A.~Tomasiello, {\it A scan for new n=1
  vacua on twisted tori},  {\em JHEP} {\bf 05} (2007) 031,
  [\href{http://xxx.lanl.gov/abs/hep-th/0609124}{{\tt hep-th/0609124}}].

\bibitem{Dall'Agata:2004dk}
G.~Dall'Agata, {\it On supersymmetric solutions of type iib supergravity with
  general fluxes},  {\em Nucl. Phys.} {\bf B695} (2004) 243--266,
  [\href{http://xxx.lanl.gov/abs/hep-th/0403220}{{\tt hep-th/0403220}}].

\bibitem{Klebanov:2000hb}
I.~R. Klebanov and M.~J. Strassler, {\it Supergravity and a confining gauge
  theory: Duality cascades and chisb-resolution of naked singularities},  {\em
  JHEP} {\bf 08} (2000) 052,
  [\href{http://xxx.lanl.gov/abs/hep-th/0007191}{{\tt hep-th/0007191}}].

\bibitem{Hailu:2006uj}
G.~Hailu and S.~H.~H. Tye, {\it Structures in the gauge / gravity duality
  cascade},  {\em JHEP} {\bf 08} (2007) 009,
  [\href{http://xxx.lanl.gov/abs/hep-th/0611353}{{\tt hep-th/0611353}}].

\bibitem{Polchinski:1998rr}
J.~Polchinski, {\em String Theory. Vol. 2: Superstring Theory and Beyond}.
\newblock Cambridge Univ. Pr., 1998.

\bibitem{Bergshoeff:2001pv}
E.~Bergshoeff, R.~Kallosh, T.~Ortin, D.~Roest, and A.~Van~Proeyen, {\it New
  formulations of d = 10 supersymmetry and d8 - o8 domain walls},  {\em Class.
  Quant. Grav.} {\bf 18} (2001) 3359--3382,
  [\href{http://xxx.lanl.gov/abs/hep-th/0103233}{{\tt hep-th/0103233}}].

\bibitem{Chiossi:2002tw}
S.~Chiossi and S.~Salamon, {\it The intrinsic torsion of $su(3)$ and $g_2$
  structures},  \href{http://xxx.lanl.gov/abs/math.dg/0202282}{{\tt
  math.dg/0202282}}.

\bibitem{Fidanza:2003zi}
S.~Fidanza, R.~Minasian, and A.~Tomasiello, {\it Mirror symmetric
  su(3)-structure manifolds with ns fluxes},  {\em Commun. Math. Phys.} {\bf
  254} (2005) 401--423, [\href{http://xxx.lanl.gov/abs/hep-th/0311122}{{\tt
  hep-th/0311122}}].

\bibitem{Hitchin:2004ut}
N.~Hitchin, {\it Generalized calabi-yau manifolds},  {\em Quart. J. Math.
  Oxford Ser.} {\bf 54} (2003) 281--308,
  [\href{http://xxx.lanl.gov/abs/math/0209099}{{\tt math/0209099}}].

\bibitem{Frey:2004rn}
A.~R. Frey, {\it Notes on su(3) structures in type iib supergravity},  {\em
  JHEP} {\bf 06} (2004) 027,
  [\href{http://xxx.lanl.gov/abs/hep-th/0404107}{{\tt hep-th/0404107}}].

\bibitem{Maldacena:2000yy}
J.~M. Maldacena and C.~Nunez, {\it Towards the large n limit of pure n = 1
  super yang mills},  {\em Phys. Rev. Lett.} {\bf 86} (2001) 588--591,
  [\href{http://xxx.lanl.gov/abs/hep-th/0008001}{{\tt hep-th/0008001}}].

\bibitem{Chamseddine:1997nm}
A.~H. Chamseddine and M.~S. Volkov, {\it Non-abelian bps monopoles in n = 4
  gauged supergravity},  {\em Phys. Rev. Lett.} {\bf 79} (1997) 3343--3346,
  [\href{http://xxx.lanl.gov/abs/hep-th/9707176}{{\tt hep-th/9707176}}].

\bibitem{Papadopoulos:2000gj}
G.~Papadopoulos and A.~A. Tseytlin, {\it Complex geometry of conifolds and
  5-brane wrapped on 2- sphere},  {\em Class. Quant. Grav.} {\bf 18} (2001)
  1333--1354, [\href{http://xxx.lanl.gov/abs/hep-th/0012034}{{\tt
  hep-th/0012034}}].

\bibitem{Aganagic:2006ex}
M.~Aganagic, C.~Beem, J.~Seo, and C.~Vafa, {\it Geometrically induced
  metastability and holography},
  \href{http://xxx.lanl.gov/abs/hep-th/0610249}{{\tt hep-th/0610249}}.

\bibitem{Burgess:2006mn}
C.~P. Burgess {\em et.~al.}, {\it Warped supersymmetry breaking},
  \href{http://xxx.lanl.gov/abs/hep-th/0610255}{{\tt hep-th/0610255}}.

\bibitem{Kallosh:2007wm}
R.~Kallosh and A.~Linde, {\it Testing string theory with cmb},  {\em JCAP} {\bf
  0704} (2007) 017, [\href{http://xxx.lanl.gov/abs/arXiv:0704.0647
  [hep-th]}{{\tt arXiv:0704.0647 [hep-th]}}].

\end{thebibliography}

\providecommand{\href}[2]{#2}\begingroup\raggedright\endgroup

\end{document}